\documentclass[letter, journal]{IEEEtran}

\usepackage{hyperref}
\pdfoutput=1
\usepackage{times,amsmath,epsfig}
\usepackage{caption}
\usepackage{algorithm}
\usepackage{algpseudocode}
\usepackage{graphicx}
\usepackage{url}
\usepackage{multirow}
\usepackage{multicol}
\usepackage{subfigure}
\usepackage{threeparttable}
\usepackage{booktabs}
\usepackage{ulem}
\usepackage{amsfonts}
\usepackage{textcomp}
\usepackage[framemethod=1]{mdframed}
\usepackage{enumitem}
\usepackage{amsthm}
\usepackage{scalerel}
\usepackage{amssymb} 
\theoremstyle{plain}
\newtheorem{theorem}{Theorem}
\theoremstyle{definition}

\theoremstyle{plain}
\newtheorem{lemma}[theorem]{Lemma}
\theoremstyle{plain}
\newtheorem{corollary}[theorem]{Corollary}
\theoremstyle{plain}
\newtheorem{prop}[theorem]{Proposition}

\ifodd 0
    \newcommand\rev[1]{{\color{blue}#1}}
    \newcommand{\com}[1]{\textbf{\color{red} (COMMENT: #1)}} 
\else
    \newcommand\rev[1]{{#1}}
    \newcommand{\com}[1]{}
\fi

\definecolor{shadecolor}{rgb}{0.878906, 0.878906, 0.878906}

\def\BibTeX{{\rm B\kern-.05em{\sc i\kern-.025em b}\kern-.08em
    T\kern-.1667em\lower.7ex\hbox{E}\kern-.125emX}}

\newcommand{\tabincell}[2]{\begin{tabular}{@{}#1@{}}#2\end{tabular}}

\title{DeepOPF: A Deep Neural Network Approach for Security-Constrained DC Optimal Power Flow}

\author{Xiang~Pan,~\IEEEmembership{Student Member,~IEEE}, Tianyu~Zhao, Minghua~Chen,~\IEEEmembership{Senior Member,~IEEE}, and Shengyu Zhang \thanks{The work presented in this paper was supported in part by a Start-up Grant (Project No. 9380118) from City University of Hong Kong. 

Xiang Pan and Tianyu Zhao are with the Department of Information Engineering, The Chinese University of Hong Kong (email: px018@ie.cuhk.edu.hk; zt017@ie.cuhk.edu.hk). Minghua Chen is with the School of Data Science, City University of Hong Kong (minghua.chen@cityu.edu.hk). Shengyu Zhang is with the Tencent Quantum Laboratory (email: shengyzhang@tencent.com). Corresponding author: Minghua Chen.
}}

\begin{document}
\begin{NoHyper}
\maketitle

\begin{abstract}
We develop \textsf{DeepOPF} as a Deep Neural Network (DNN) approach for solving security-constrained direct current optimal power flow (SC-DCOPF) problems, which are critical for reliable and cost-effective power system operation. \textsf{DeepOPF} is inspired by the observation that solving SC-DCOPF problems for a given power network is equivalent to depicting a high-dimensional mapping from the load inputs to the generation and phase angle outputs. We first train a DNN to learn the mapping and predict the generations from the load inputs. We then directly reconstruct the phase angles from the generations and loads by using the power flow equations. Such a predict-and-reconstruct approach 
reduces the dimension of the mapping to learn, subsequently cutting down the size of the DNN and the amount of training data needed. We further derive a condition for tuning the size of the DNN according to the desired approximation accuracy of the load-generation mapping. {We develop a post-processing procedure based on $\ell_1$-projection to ensure the feasibility of the obtained solution, which can be of independent interest. Simulation results for IEEE test cases show that \textsf{DeepOPF} generates feasible solutions with less than \rev{0.2\%} optimality loss, while speeding up the computation time by up to two orders of magnitude as compared to a state-of-the-art solver.}  
\end{abstract}
\begin{IEEEkeywords}
Deep learning, Deep neural network, Optimal power flow.
\end{IEEEkeywords}
\vspace*{-2mm}
\section*{Nomenclature}
{\small
\begin{IEEEdescription}[\IEEEusemathlabelsep\IEEEsetlabelwidth{MMM}]
    \item[Variable\ \ Definition]
    \item[$\mathcal{N}$] Set of buses, $N \triangleq |\mathcal{N}|$
    \item[$\mathcal{E}$] Set of branch 
    \item[$\mathcal{G}$] Set of generators 
    \item[$\mathcal{D}$] Set of load 
    \item[$\mathcal{C}$] Set of contingency cases 
    \item[$P_G$] Power generation injection vector, $[P_{G_i}, i\in \mathcal{N}]$
    \item[$P_G^{\min}$] Minimum generator output vector, $[P^{\min}_{G_i}, i\in \mathcal{N}]$
    \item[$P_G^{\max}$] Maximum generator output vector, $[P^{\max}_{G_i}, i\in \mathcal{N}]$
    \item[$P_D$] Power load vector, $[P_{D_i}, i\in \mathcal{N}]$
    \item[$\varTheta_c$] Voltage angle vector under the $c$-th contingency
    \item[$\theta_{c,i}$] Voltage angle under the $c$-th contingency for bus $i$
    \item[$\mathbf{B}_c$] Admittance matrix under the $c$-th contingency
    \item[${x_{ij,c}}$] Line reactance from bus $i$ to $j$ under the $c$-th contingency
    \item[$P_{Tij,c}^{\max}$] Line transmission limit from bus $i$ to $j$ under the $c$-th contingency
    \item[$N_{\scaleto{hid}{3pt}}$] The number of hidden layers in the neural network
\end{IEEEdescription}
We use $|\cdot|$ to denote the size of a set. $P_{G_i}=P^{\min}_{G_i}=P^{\max}_{G_i}=0, \forall i \notin \mathcal{G}$, and $P_{D_i}=0, \forall i \notin \mathcal{D}$.  
}

\vspace*{-5mm}

\section{Introduction}\label{sec:intro}
The ``deep learning revolution'' largely enlightened by the October 2012 ImageNet victory~\cite{krizhevsky2012imagenet} has transformed various industries in human society, including artificial intelligence, health care, online advertising, transportation, and robotics. As the most widely-used and mature model in deep learning, Deep Neural Network (DNN) ~\cite{goodfellow2016deepma}  demonstrates superb performance in complex engineering tasks such as recommendation ~\cite{covington2016deep}, bio-informatics ~\cite{bty543}, mastering difficult game like Go ~\cite{silver2016mastering}, and human pose estimation ~\cite{6909610}. 
The capability of approximating continuous mappings and the desirable scalability make DNN a favorable choice in the arsenal of solving large-scale optimization and decision problems in engineering systems.  In this paper, we apply DNN to power systems for solving the essential security-constrained direct current optimal power flow (SC-DCOPF) problem in power system operation.

The OPF problem, first posed by Carpentier in 1962 in~\cite{carpentier1962contribution}, is to minimize an objective function, such as the cost of power generation, subject to all physical, operational, and technical constraints, by optimizing the dispatch and transmission decisions. These constraints include Kirchhoff's laws, operating limits of generators, voltage levels, and loading limits of transmission lines~\cite{johnson1989electric}. The OPF problem is central to power system operations as it underpins various applications including economic dispatch, unit commitment, stability and reliability assessment, and demand response. While OPF with a full AC power flow formulation (AC-OPF) is most accurate, it is a non-convex problem and its complexity obscures practicability. Meanwhile, based on linearized power flows, DC-OPF is a convex problem admitting a wide variety of applications, including electricity market clearing and power transmission management. See e.g.,~\cite{frank2012optimal1, frank2012optimal2} for a survey.

The SC-DCOPF problem, a variant of DC-OPF, is critical for reliable power system operation against contingencies caused by equipment failure~\cite{cain2012history}. It considers not only constraints under normal operation, but also additional steady-state security constraints for each possible contingency\footnote{There are two types of SC-DCOPF problems, namely the preventive SC-DCOPF problem and the corrective SC-DCOPF problem. Both of them are critical in practice. We focus on the preventive SC-DCOPF problem in this paper, in which the system operating decisions stay unchanged once determined and they need to satisfy both the pre- and post- contingency constraints. Usually, only line contingencies are considered in the preventive SC-DCOPF problem~\cite{ardakani2013identification}. Our \textsf{DeepOPF} approach is also useful for the corrective SC-DCOPF problem, where the system operator only has a short time to adjust the operating points after the occurrence of a contingency. By \textsf{DeepOPF}, the system operator can obtain new operating points in a fraction of the time used by conventional solvers.}~\cite{CAPITANESCU20111731}. Meanwhile, solving SC-DCOPF incurs excessive computational complexity, limiting its applicability in large-scale power networks~\cite{chiang2015solving}.

To this end, we propose a machine learning approach for directly solving the SC-DCOPF problem. Our approach is based on the following observations. 
\begin{itemize}
    \item Given a power network, solving the SC-DCOPF problem is equivalent to depicting a high-dimensional mapping between load inputs and generations and voltages outputs.
    
    \item In practice, the SC-DCOPF problem is usually solved repeatedly for the same network, e.g., every 5 minutes~\cite{cain2012history}, with different load inputs at different time epochs. 
\end{itemize} 
As such, it is conceivable to leverage the universal approximation capability of deep feed-forward neural networks~\cite{leshno1993multilayer,hanin2017approximating,kidger2019universal,hornik1991approximation,karg2018efficient}, to learn the input-to-output mapping for a given power network, and then apply the mapping to obtain operating decisions upon giving load inputs (e.g., once every 5 minutes).\footnote{{Given a power network, as discussed in Sec.~\ref{sec:OPF.theory_analysis2}, the mapping between the load input and the optimal solution of the SC-DCOPF problem is continuous and piece-wise linear. Existing works~\cite{hornik1991approximation,leshno1993multilayer,hanin2017approximating,kidger2019universal,karg2018efficient} show that the feed-forward neural networks can approximate real-valued continuous functions arbitrary well as the neural network size goes to infinity. Thus one can expect that a well-trained DNN would generate a close-to-optimal solution for the SC-DCOPF problem.}} 

Specifically, we develop \textsf{DeepOPF} as a DNN based solution for the SC-DCOPF problem. As compared to conventional approaches based on interior-point methods~\cite{ye1989extension}, \textsf{DeepOPF} excels in (i) reducing computing time and (ii) scaling well with the problem size. These salient features are particularly appealing for solving large-scale SC-DCOPF problems. Note that the complexity of constructing and training a DNN model is minor if amortized over many problem instances (e.g., one per every 5 minutes) that can be solved using the same model. We summarize our contributions as follows.

First, after reviewing the SC-DCOPF problem in Sec.~\ref{sec:OPF.review}, we prospose \textsf{DeepOPF} as a DNN framework for solving the SC-DCOPF problem in Sec.~\ref{sec:DeepOPF}. In \textsf{DeepOPF},  we first train a DNN to learn the load-generation mapping and predict the generations from the load inputs. We then directly reconstruct the phase angles from the generations and loads by using the (linearized) power flow equations. Such a predict-and-reconstruct two-step procedure reduces the dimension of the mapping to learn, subsequently cutting down the size of our DNN and the amount of training data/time needed. {We also design {a post-processing procedure based on $\ell_1$-projection} to ensure the feasibility of the final solution, which can be of independent interest.} 

Then in Sec.~\ref{sec:OPF.theory_analysis2}, we derive a condition suggesting that the approximation accuracy of the neural network in \textsf{DeepOPF} decreases exponentially in the number of layers and polynomially in the number of neurons per layer. This allows us to tune the size of the neural network in \textsf{DeepOPF} according to the desired performance. We also analyze the computational complexity of \textsf{DeepOPF}.

Finally, we carry out simulations and summarize the results in Sec.~\ref{sec:simulations}. {Simulation results of IEEE test cases show that \textsf{DeepOPF} generates feasible solutions with less than \rev{0.2\%} optimality loss. As compared to a state-of-the-art solver, \textsf{DeepOPF} speeds up the computation time by up to two orders of magnitude under the typical load condition and by up to one order of magnitude under the congested load condition. The results also suggest a trade-off between the prediction accuracy and running time of \textsf{DeepOPF}.}

Due to the space limitation, all proofs are in the supplementary material. 
\section{Related Work}\label{sec:OPF.relatedwork}
Existing studies on solving SC-OPF focus on four lines of approaches. The first is on iteration-based algorithms. The SC-OPF problem is first approximated as an optimization problem, e.g., quadratic programming~\cite{71294} or linear programming~\cite{6756976}. Then iteration-based algorithms, e.g., the interior-point method~\cite{ye1989extension,wang2007computational}, are applied to solve the approximated problems. The time complexity of iteration-based algorithms, however, can be substantial for large-scale power systems, limiting its applicability in practice. This is due to the significant number of constraints introduced by the consideration of a large number of contingencies. See, e.g.,~\cite{CAPITANESCU20111731} for a survey on the iteration-based algorithms for solving SC-OPF problems. 

The second approach is based on computational intelligence, e.g., evolutionary programming~\cite{yumbla2008optimal,amjady2012solution,capitanescu2016critical}. For instance, the authors of~\cite{yumbla2008optimal} propose a particle swarm optimization method for solving SC-OPF problems, in which they apply the particle swarm optimization (PSO) algorithm with reconstruction operators (PSO-RO) to find the solutions and designed an external penalty to ensure the feasibility of the obtained solution. 
Two limitations of this approach are the lack of performance guarantee and high computational complexity~\cite{vikhar2016evolutionary}. 

The third is on learning-based methods. There have been researches applying machine learning to various tasks in the power system, e.g., power system state estimation (PSSE)~\cite{zhang2019real}; see~\cite{wehenkel2012automatic} for a comprehensive survey. On solving OPF problems, existing studies mainly focus on integrating the learning techniques into conventional algorithms to facilitate the solving process~\cite{gutierrez2010neural,hasan2020survey}. 
For instance,~\cite{gutierrez2010neural} applies a neural network to learn the system security boundaries as an explicit function to be used in the OPF formulation.

Recently, there is a line of research on determining the active/inactive constraints set to reduce the size of power system optimization problems, e.g., unit commitment and  OPF problems, to accelerate the solving process~\cite{zhai2010fast, roald2019implied, 9091534}. The idea of this category of approaches is to reduce the scale of the problem without losing optimality. The speedup comes from the problem size reduction as any solvers can solve the reduced problem faster than solving the original problem directly. There is no optimality loss for this category of approach if the inactive/active constraints are identified correctly. Meanwhile, our \textsf{DeepOPF} approach develops a DNN-based solver for the OPF problem. It relies on having a large number of training data to train a DNN to predict generations from the input loads. The approach designs one solver for every interested OPF formulation. The advantage lies in that the DNN solvers can be much faster than conventional solvers. The optimality loss can be adjusted by constructing and training a larger DNN. The speedup comes from the new framework for designing OPF solvers. The disadvantage is two-fold. First, the approach would always incur optimality loss, due to the approximation errors of DNN. Second, one will need to design and train different solvers for different power networks and prepare a large amount of training data. In practice, these can be done offline. 

We note that two approaches are orthogonal to each other and can be combined to achieve a speedup better than those by individual approaches. Specifically, given an OPF formulation, one can first identify and remove the inactive constraints to obtain a reduced problem (assuming no optimality loss). Then one can apply the pre-trained DNN solver for the reduced problem to obtain the solution. This way, the overall speedup performance is the product of that achieved by reducing problem size and that achieved by the DNN solver, better than those achievable by individual approaches. Both approaches may generate infeasible solutions. In such cases, one may apply the $\ell_1$-projection based post-processing procedure, in Sec.~\ref{ssec:post.processing}, to obtain a feasible solution with less computational time than re-solving the original (quadratic) SC-DCOPF problem.


It is also conceivable to apply the K nearest neighbor (KNN) scheme to generate an approximate solution of the OPF problems given the load input. It is well understood that, as compared to the neural-network based approach, the KNN scheme incurs less training-time complexity but higher running-time complexity. This is due to that one has to identify the K nearest neighbors of the input, which is expensive for large-dimension problems. 

To our best knowledge, \textsf{DeepOPF} is the first to develop a DNN-based solver for directly solving OPF problems. It learns the mapping from the load inputs to the generation and voltage outputs and directly obtains solutions for the SC-DCOPF problem with feasibility guarantees. As compared to our previous study in~\cite{pan-zhao-chen.2019}, this paper studies the more challenging SC-DCOPF problem and characterizes a condition for tuning the size of the DNN according to the desired approximation  accuracy  of  the  load-generation mapping. The predict-and-reconstruct DNN framework for solving OPF problems outlined in~\cite{pan-zhao-chen.2019} (and this paper) applies to the AC-OPF setting as well. It has received growing interests with initial results reported in \cite{pan2020deepopf,zamzam2019learning}, which demonstrate the speedup potential and highlight the challenges of ensuring solution feasibility. 
\begin{figure}[!t]
	\centering
	\includegraphics[width = 0.5\textwidth]{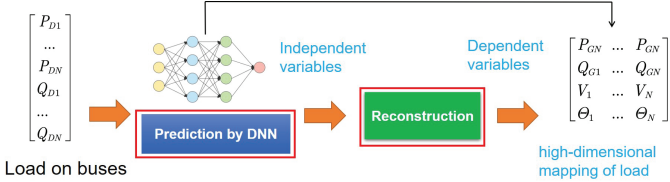}
	\caption{Overview of the predict-and-reconstruct framework.} 
	\label{fig1}
\end{figure}

\section{Security-Constrained DCOPF problem} \label{sec:OPF.review}
We focus on the widely-studied $(N-1)$ SC-DCOPF problem considering contingencies due to the outage of any single transmission line. 
The objective is to minimize the total generation cost subject to the generator operation limits, the power balance equations, and the transmission line capacity constraints under all contingencies~\cite{4956966}. 
Assumed that the power network remains connected upon contingency, the SC-DCOPF problem is formulated as follows\footnote{We note that there is another formulation involving only generations as the phase angels can be uniquely determined by the generations and loads; see e.g.,~\cite{4956966}. We focus on the standard formulation and both formulations incur the same order of running time complexity~\cite{en11061497}.}:
\begin{align}
    \min_{\varTheta_c, P_G} \quad &\mathrm{\ }\sum_{i\in\mathcal{G}}^{}{g_i\left( P_{Gi} \right)} \label{eq:SC-DCOPF.obj}\\
    \mathrm{s.t.} \quad& P_{Gi}^{\min}\le P_{Gi}\le P_{Gi}^{\max},\,\,i\in \mathcal{G}, \label{eq:SC-DCOPF.generator.limit}\\
    &  \mathbf{B}_c\cdot \varTheta_c=P_G-P_D, c\in \mathcal{C}, \label{eq:SC-DCOPF.pf}\\
    & \frac{1}{x_{ij,c}}\left( \theta _{i,c}-\theta _{j,c} \right) \le P_{Tij,c}^{\max},\,\,(i,j)\in \mathcal{E}, c\in \mathcal{C}. 
    \label{eq:SC-DCOPF.line.capacity}
\end{align}
Here $c=0$ denotes the case without any contingencies. $P_{Tij,c}^{\max}$ is the transmission limit for the branch connecting buses $i$ and $j$.
$\mathbf{B}_c$ is the admittance matrix for the $c$-th contingency, which is an $N \times N$ matrix with entries
\noindent
\[ B_{ij,c} =
  \begin{cases}
  0, & \quad \mbox{if } \ (i, j)\notin \mathcal{E}, i\neq j;\\
    -\frac{1}{x_{ij,c}},      & \quad \mbox{if } \ (i, j)\in \mathcal{E};\\
    \displaystyle\sum_{k=1,k\neq i}^{N} \frac{1}{x_{ij,c}}, & \quad \mbox{if} \quad i=j.
  \end{cases}
\]
The first set of constraints in the formulation describe the generation limits. The second set of constraints are the power flow equations with contingencies taken into account. The third set of constraints capture the line transmission capacity for both pre-contingency and post-contingency cases. In the objective, $g_i\left( P_{Gi} \right)$ is the cost function for the generator at the $i$-th bus, commonly modeled as a quadratic function~\cite{260897}: 
\begin{equation}
g_i\left(P_{Gi} \right) =\lambda _{1i}P_{Gi}^2+\lambda _{2i}P_{Gi}+\lambda _{3i}, \label{equation3}
\end{equation}
where $\lambda_{1i}$, $\lambda_{2i}$, and $\lambda_{3i}$ are the model parameters and can be obtained from measured data of the heat rate curve~\cite{823997}. We note that the SC-DCOPF problem is a strictly convex (quadratic) problem and thus has a unique optimal solution. While the SC-DCOPF problem is important for reliable power system operation, solving it for large-scale power networks incurs excessive running time, limiting its practicability~\cite{chiang2015solving}. In the next section, we propose a neural network approach to solve the SC-DCOPF problem in a fraction of the time used by existing solvers.

\section{{\textsf{DeepOPF}} for Solving SC-DCOPF} \label{sec:DeepOPF}
\subsection{A Neural-Network Approach for Solving OPF Problems}\label{ssec:predict-and-reconstrcut.framework}
We outline a general predict-and-reconstruct framework for solving OPF in Fig.~\ref{fig1}. Specifically, we exploit the dependency induced by the equality constraints among the decision variables in the OPF formulation. Given the load inputs, the learning model (e.g., DNN) is applied to predict only a set of \textit{independent} variables. The remaining variables are then determined by leveraging the (power balance) equality constraints. This way, we not only reduce the number of variables to predict but also guarantee that the obtained solution always satisfies the equality constraints, which is usually difficult for generic learning based approaches. In this paper, we follow this general approach to develop \textsf{DeepOPF} for solving the SC-DCOPF problem. 

\begin{figure*}[!t]
	\centering
	\includegraphics[width = 0.7\textwidth]{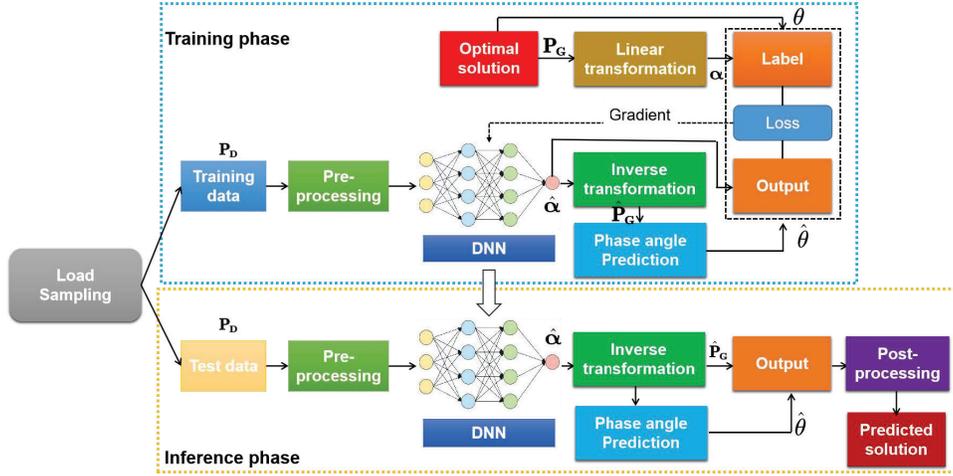}
	\caption{The flow chart of \textsf{DeepOPF}.}
	\label{fig3}
\end{figure*}

\subsection{Overview of \textsf{DeepOPF}}
The framework of \textsf{DeepOPF} is shown in Fig.~\ref{fig3}, which is divided into a training stage and an inference stage. 
We first train a DNN to learn the load-generation mapping and predict the generations from the load inputs. We then directly compute the voltages from the generations and loads by using the (linearized) power flow equations. 

We discuss the process of constructing and training the DNN model in the following subsections. In particular, we discuss the preparation of the training in Sec.~\ref{ssec:load.sampling.and.pre-preprocessing}, the variable prediction and reconstruction in Sec.~\ref{ssec:linear.transformation.and.dimension.reduction}, and the design and training of DNN in Sec.~\ref{ssec:DNN.and.loss.function}. 

In the inference stage, we directly apply \textsf{DeepOPF} to solve the SC-DCOPF problem with given load inputs. {\textsf{DeepOPF} may generate infeasible solutions due to the error in approximating the mapping. We describe a post-processing procedure based on $\ell_1$-projection to ensure the feasibility of the obtained solutions in Sec.~\ref{ssec:post.processing}}.

\subsection{Load Sampling and Pre-processing}\label{ssec:load.sampling.and.pre-preprocessing}
We sample the loads within $[(1-x)\cdot P_{Di}, (1+x)\cdot P_{Di}]$ uniformly at random, where $P_{Di}$ is the default power load at the $i$-th bus and $x$ is the percentage of sampling range, e.g., 10$\%$. It is then fed into the traditional quadratic programming solver~\cite{gurobi} to generate the optimal solutions. Uniform sampling is applied to avoid the over-fitting issue which is common in generic DNN approaches\footnote{For load inputs of large dimensions, the uniform mechanism may not be sufficient to guarantee enough good samples, especially near the boundary. In those cases, Markov chain Monte Carlo (MCMC) methods can be applied to sample according to a pre-specified probability distribution, to collect sufficient samples near the boundary of the sampling space.}. After that, the training data is normalized (using the statistical mean and standard variation) to improve training efficiency. 

\subsection{Generation Prediction and Phase Angle Reconstruction} \label{ssec:linear.transformation.and.dimension.reduction}
We express $P_{Gi}$ as follows, 
for $i\in \mathcal{G}$,
\begin{equation}
P_{Gi}=\alpha _i\cdot\left( P_{Gi}^{\max}-P_{Gi}^{\min} \right)+P_{Gi}^{\min}, \label{equation4}
\end{equation}
where $\alpha _i\in \left[ 0,1 \right]$ is a scaling factor. It is clear that $\alpha_i$ and $P_{G_i}$ have a one-to-one correspondence. Thus the scaling factors used in the training phase can be directly computed from the generated data. Meanwhile, instead of predicting the generations with diverse value ranges, we predict the scaling factor $\alpha_i\in[0,1]$ and recover $P_{Gi}$ by using (\ref{equation4})). This simplifies the DNN output layer design to be discussed later. Note that the generation of the slack bus is obtained by subtracting generations of other buses from the total load.

Once we obtain $P_G$, we directly compute the phase angles by a useful property of the admittance matrices. We first obtain an $(N-1) \times (N-1)$ matrix, $\mathbf{\tilde{B}}_c$ by eliminating the row and column corresponding to the slack bus from the admittance matrix $\mathbf{B}_c$ for the $c$-th contingency. It is well-understood that $\mathbf{\tilde{B}}_c$ is a full-rank matrix ~\cite{823997}. Then we compute an {$(N-1)$-dimensional} phase angle vector $\tilde{\varTheta}_c$ as 
\begin{equation}
    \tilde{\varTheta}_c=\left( \mathbf{\tilde{B}}_c \right) ^{-1}\left( \tilde{P}_G-\tilde{P}_D \right),
    \label{equation5} 
\end{equation}
where $\tilde{P}_G$ and $\tilde{P}_D$ stand for the {$(N-1)$-dimensional} generation and load vectors for buses excluding the slack bus under each contingency, respectively.
In the end, we output the {$N$-dimensional} phase angle vector $\varTheta_c$ by inserting a constant phase angle for the slack bus into $\tilde{\varTheta}_c$.

There are two advantages to this design. On one hand, we use the property of the admittance matrix to reduce the number of variables to predict by our neural network, cutting down the size of our DNN model and the amount of training data/time needed. On the other hand, the equality constraints involving the generations and the phase angles can be satisfied automatically, which can be difficult to handle in alternative learning-based approaches. 

\subsection{The DNN Model} \label{ssec:DNN.and.loss.function}
The core of \textsf{DeepOPF} is the DNN model, which is applied to approximate the load-generation mapping, given a power network. The DNN model is established based on the multi-layer feed-forward neural network structure, which consists of typical three-level network architecture: one input layer, several hidden layers, and one output layer. More specifically, the DNN model is defined as:
\begin{eqnarray*}
    &h_0&=P_D,\\
    &h_i&=\sigma \left( W_ih_{i-1}+b_{i-1} \right), \forall \ i=1, ..., N_{\scaleto{hid}{3pt}}\\ 
    &\hat{\alpha}&=\sigma '\left( w_oh_{\scaleto{hid}{3pt}}+b_o \right), 
\end{eqnarray*}
where $h_0$ denotes the input vector of the network, $h_i$ is the output vector of the $i$-th hidden layer and $\hat{\alpha}$ is the generated scaling factor vector for the generators.

\subsubsection{The architecture}
The $i$-th hidden layer models the interactions between features by introducing a connection weight matrix $W_i$ and a bias vector $b_i$. The activation function $\sigma(\cdot)$ further introduces non-linearity into the hidden layers. We adopt the Rectified Linear Unit (ReLU) as the activation function of the hidden layers, which helps to accelerate the convergence and alleviate the vanishing gradient problem~\cite{krizhevsky2012imagenet}. In addition, the Sigmoid function~\cite{goodfellow2016deepma}, $\sigma '\left( x \right) =\frac{1}{1+e^{-x}}$, is applied on the output layer to constrain the outputs within $(0, 1)$. 


\subsubsection{The loss function}
After constructing the DNN model, we need to design the corresponding loss function to guide the training. Since there exists a one-to-one correspondence between $P_G$ and $\varTheta_c$, it suffices to focus on the loss of $P_G$, which is defined as the sum of mean square error between the obtained $\hat{\alpha}_i$ and the optimal scaling factors ${\alpha}_i$ as follows:
\begin{eqnarray}
\mathcal{L}_{P_G}=\frac{1}{|\mathcal{G}|}\sum_{i\in\mathcal{G}}^{}{\left( \hat{\alpha}_i-\alpha _i \right) ^2}.
\label{equation10}
\end{eqnarray}

Meanwhile, we introduce a penalty term related to the inequality constraint into the loss function. We first introduce an $N_a \times N$ matrix $\mathbf{A}_c$ for each contingency $c$, where $N_a$ is the number of adjacent buses. Each row in $\mathbf{A}_c$ corresponds to an adjacent bus pair. Given the $k$-th adjacent bus pair $(i_k,j_k)\in\mathcal{E}$, $k=1,...,N_a$, under the $c$-th contingency, let the power flow from the $i_k$-th bus to the $j_k$-th bus. Thus, the elements, $a_{ki_k,c}$ and $a_{kj_k,c}$, the corresponding entries of the matrix $\mathbf{A}_c$, are given as:
\begin{equation}
    a_{ki_k,c}=\frac{1}{P_{Ti_kj_k,c}^{\max}\cdot x_{i_kj_k,c}}\ \mathrm{and\ }a_{kj_k,c}=\frac{-1}{P_{Ti_kj_k,c}^{\max}\cdot x_{i_kj_k,c}}.
\label{equation8}
\end{equation}
Based on (\ref{equation5}) and (\ref{equation8}), the capacity constraints for the transmission line in \eqref{eq:SC-DCOPF.line.capacity} can be expressed as:
\begin{equation}
-1\le \left(\mathbf{A}_c \hat{\varTheta}_c \right)_k\le 1, k=1,...,N_a, c\in \mathcal{C},
\label{equation9}
\end{equation}
\noindent where $(\mathbf{A}_c \hat{\varTheta}_c )_k$ represents the $k$-th element of $\mathbf{A}_c \hat{\varTheta_c}$. Note that $\hat{\varTheta}_c$ is the phase angle vector generated based on \eqref{equation5} and the discussion below it, and it is computed from $P_G$ and $P_D$. We can then calculate  $(\mathbf{A}_c\hat{\varTheta}_c )_k$. The penalty term capturing the feasibility of the generated solutions is defined as:
\begin{eqnarray}
\mathcal{L}_{pen}=\frac{1}{N_a}\sum_{k=1}^{N_a}{\max \left( \left( \mathbf{A}_c\hat{\varTheta}_c \right) _{k}^{2}-1,0 \right) }.
\label{equation12}
\end{eqnarray}
In summary, the loss function consists of two parts: the difference between the generated solution and the reference solution and the penalty upon solutions violating the inequality constraints. The total loss is a weighted sum of the two:
\begin{eqnarray}
\mathcal{L}_{total}=w_1\cdot \mathcal{L}_{P_G}+w_2\cdot \mathcal{L}_{pen},
\label{equation13}
\end{eqnarray}
\noindent where $w_1$ and $w_2$ are positive weighting factors for balancing the influence of each term in the training phase. 

\subsubsection{The training process}
The training processing can be regarded as minimizing the average loss for the given training data by tuning the parameters of the DNN model as follows:
\begin{equation}\label{equation11}
    \min_{W_i, b_i }\frac{1}{N_{\scaleto{T}{3pt}}}\sum_{k=1}^{N_{\scaleto{T}{3pt}}}{\mathcal{L}_{total,k}}
\end{equation}
\noindent where we recall that $W_i$ and $b_i$, $i=1,...,N_{\scaleto{hid}{3pt}}$ represent the connection weight matrix and vector for layer $i$. $N_{\scaleto{T}{3pt}}$ is the amount of training data and $\mathcal{L}_{total,k}$ is the loss of the $k$-th item in the training. 

We apply the stochastic gradient descent (SGD) method with momentum~\cite{qian1999momentum} to solve the problem in (\ref{equation11}), which is effective for the large-scale dataset and can economize on the computational cost at every iteration by choosing a subset of summation functions at every step. 

\subsection{Post-Processing} \label{ssec:post.processing}
After obtaining a solution including the generations and phase angles, we check its feasibility by examining if it violates the generation limits and the line transmission limits. We output the solution if it passes the feasibility test. Otherwise, we solve the following $\ell_1$-projection problem with linear constraints to obtain a feasible solution, \footnote{{It is common for machine learning approaches to generate infeasible solutions. The proposed post-processing procedure can then be applied to recover a feasible solution. The simulation results in Sec.~\ref{sec:simulations} show that \textsf{DeepOPF} with post-processing achieves decent speedup performance.}}
\begin{equation}
    {\min}\ \lVert \hat{P}_G-U \rVert_{1}
    \mathrm{\ \ s.t.\ \ }U\ \text{satisfies \eqref{eq:SC-DCOPF.generator.limit}-\eqref{eq:SC-DCOPF.line.capacity},}
    \label{equation14}
\end{equation}
where $\hat{P}_G$ is the solution predicted by DNN. We remark that such an $\ell_1$-projection problem is indeed an LP and can be solved by off-the-shell solvers.

\section{Performance Analysis of \textsf{DeepOPF}}\label{sec:OPF.theory_analysis2}
\subsection{Approximation Error of the Load-to-Generation Mapping}\label{sub_approximation_error}
Given a power network, the SC-DCOPF problem is a quadratic programming problem with linear constraints. We denote the mapping between the load input $P_D$ and the optimal generation $P_G$ as $f^*(\cdot)$. Following the common practice in the deep-learning analysis (e.g.,~\cite{yarotsky2017error,Safran2017,liang2016deep}) and without loss of generality, we focus on the case of {one-dimensional} output in the following analysis, i.e., $f^*(\cdot)$ is a scalar.\footnote{To extend the results for mappings with {one-dimensional} output to mappings with {multi-dimensional} outputs, one can view the latter as multiple mappings each with {one-dimensional} output, apply the results for {one-dimensional} output multiple times, and combine them to get the one for {multi-dimension} output.} 
Assumed the load input domain is compact, which usually holds in practice, $f^*(\cdot)$ has certain properties.
\begin{lemma}
    The function $f^*(\cdot)$ is piece-wise linear and Lipschitz-continuous. That is, there exists a constant $\varLambda>0$, such that for any $x_1, x_2$ in the domain of $f^*(\cdot)$,
    \[
        \left|f^*(x_2)-f^*(x_1)\right| \leq \varLambda \cdot \lVert x_1-x_2 \rVert_2.
    \]
    \label{lem1}
\end{lemma}
Define $f(\cdot)$ as the mapping between $P_D$ and the generation obtained by \textsf{DeepOPF} by using a neural network with depth $N_{\scaleto{hid}{3pt}}$ and maximum number of neurons per layer $M$. We focus on the case of {one-dimensional} output. As $f(\cdot)$ is generated from a neural network with ReLU activation functions, it is also piece-wise linear~\cite{NIPS2014_5422}.

By exploiting the piece-wise linearity and the Lipschitz continuity, we analyze the approximation error between $f^*(\cdot)$ and $f(\cdot)$. 
\begin{theorem}\label{thm:worst-cast-approximation-error}
Let $\mathcal{H}$ be the class of all possible $f^*(\cdot)$ with a Lipschitz constant $\varLambda>0$. Let $\mathcal{K}$ be the class of all $f(\cdot)$ generated by a neural network with depth $N_{\scaleto{hid}{3pt}}$ and at most $M$ neurons per layer. 
\begin{equation}
\underset{f^*\in \mathcal{H}}{\max}\ \underset{f\in \mathcal{K}}{\min} \,\underset{x\in \mathcal{S}}{\max}\left| f^*\left( x \right) -f\left( x \right) \right|\geq \varLambda\cdot \frac{ d}{4\cdot(2M)^{N_{\scaleto{hid}{3pt}}}},
\label{err_bound_dnn}
\end{equation}
where $d$ is the diameter of the load input domain $\mathcal{S}$.
\end{theorem}

The theorem characterizes a lower bound on the worst-case error of using neural networks to approximate load-generation mappings in SC-DCOPF problems. The bound is linear in $d$, which captures the size of the load input domain, and $\varLambda$, which captures the ``curveness'' of the mapping to learn. Meanwhile, interestingly, the bound decreases exponentially in the number of layers while polynomially in the number of neurons per layer. This suggests the benefits of using ``deep'' neural networks in mapping approximation, similar to the observations in~\cite{yarotsky2017error,Safran2017,liang2016deep}\footnote{While our observations are similar to those in~\cite{yarotsky2017error,Safran2017,liang2016deep}, there is distinct difference in the results and the proof techniques as we explore the piece-wise linearity of the function unique to our setting.}.

A useful corollary suggested by Theorem~\ref{thm:worst-cast-approximation-error} is the following.
\begin{corollary}
\label{col1}
The following gives a condition on the neural network parameters, such that it is ever possible to approximate the most difficult load-to-generation mapping with a Lipschitz constant $\varLambda$, up to an error of $\epsilon>0$.
\begin{equation}
(2M)^{N_{\scaleto{hid}{3pt}}}\geq \varLambda \cdot \frac{d}{4\cdot \epsilon},
\label{err_bound_dnn2}
\end{equation}
where $d$ is the diameter of the input domain $\mathcal{S}$.
\end{corollary}

The condition in \eqref{err_bound_dnn2} gives a necessary ``size'' of the neural network to achieve preferred approximation accuracy. If \eqref{err_bound_dnn2} is not satisfied, then there may exist a difficult mapping, even the smallest possible approximation error exceeds $\epsilon$.

\subsection{Computational Complexity} \label{ssec:comp.complexity}
Recall that $N$ is the number of buses. The number of optimization variables in SC-DCOPF, including the generations and the phase angles of all the lines under all possible contingencies, and the constraints is $\mathcal{O}\left( N^3\right)$. 

The computational complexity of interior point methods for solving SC-DCOPF as a convex quadratic problem is $\mathcal{O} \left( \left( N^{3}\right)^4\right)=\mathcal{O} \left( N^{12}\right)$, measured as the number of elementary  operations assuming that each elementary operation takes a fixed amount of time to perform~\cite{ye1989extension}. 

The computational complexity of \textsf{DeepOPF} consists of three parts. The first is the complexity of predicting the generations using the DNN, which is $\mathcal{O} \left(N_{\scaleto{hid}{3pt}}M^2\right)$ where $M$ is the maximum number of neurons in each layer and $N_{\scaleto{hid}{3pt}}$ is the number of hidden layers in DNN. See Appendix~\ref{apx:Complexity_DNN_prediction} of the supplementary materials for details of the analysis. To achieve satisfactory performance in terms of optmality loss and speed-up, we set $M$ to be $\mathcal{O} \left(N\right)$ and $N_{\scaleto{hid}{3pt}}$ to be 3. As such, the complexity for predicting the generations by our DNN is $\mathcal{O} \left(N^2\right)$.
    
The second is the complexity of computing the phase angles from the generations by directly solving (linearized) power flow equations and checking the feasibility of the results. The process involves solving $\mathcal{O}\left(N^2 \right)$ sets of linear equations, one set for each contingency, and checking the transmission line limit constraints. The total complexity is $\mathcal{O}\left(N^5 \right)$. 

The third is the complexity of $\ell_1$-projection, if the post-processing procedure is involved to ensure feasibility of the obtained solutions. The $\ell_1$-projection is a linear programming problem and can be solved in $\mathcal{O} \left( \left( N^{3}\right)^{2.5}\right)=\mathcal{O} \left( N^{7.5}\right)$ amount of time by using algorithms based on fast matrix multiplication.


{Overall, the total computational complexity of \textsf{DeepOPF} is $\mathcal{O}\left(N^5 \right)$ if the post-processing procedure is not involved, for example, when the power system is operated in the light-load regime. Otherwise, it is $\mathcal{O} \left( N^{7.5}\right)$. In both cases, the complexity is substantially lower than that of solving the original SC-DCOPF problem directly by the conventional interior point method, which is $\mathcal{O} \left( N^{12}\right)$. 

Our simulation results in Sec.~\ref{sec:simulations} corroborate the above observations. For both typical and {congested} settings, \textsf{DeepOPF} obtains quality solutions for SC-DCOPF problems in a fraction of the time used by a state-of-the-art solver with less than 0.2\% optimality loss. {We also note that the $\ell_1$-projection in the post-processing procedure is an LP and can be solved efficiently by off-the-shelf solvers.} }

\subsection{Trade-off between Accuracy and Complexity}
The results in Theorem~\ref{thm:worst-cast-approximation-error} and Proposition~\ref{prop:DeepOPF.complexity} suggest a trade-off between accuracy and complexity. In particular, we can tune the number of hidden layers $N_{\scaleto{hid}{3pt}}$ and the maximum number of neurons per layer $M$ to trade between the approximation accuracy and computational complexity of the DNN approach. It appears desirable to design multi-layer neural networks in \textsf{DeepOPF} as increasing $N_{\scaleto{hid}{3pt}}$ may reduce the approximation error exponentially, but only increase the complexity linearly.

\begin{figure}[!t]
	\centering
	\includegraphics[width = 0.5\textwidth]{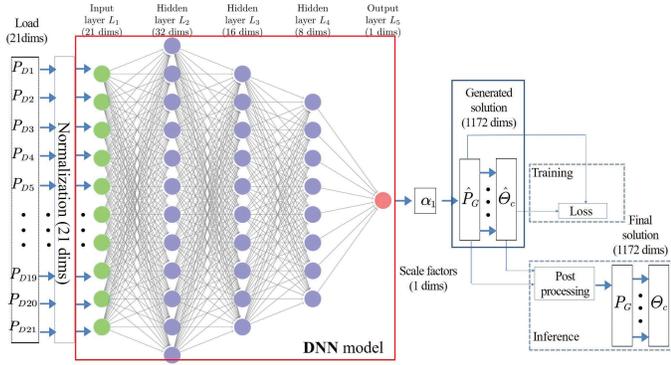}
	\caption{The detail architecture of DNN model for IEEE Case30.}
	\label{fig4}
\end{figure}

\begin{table}[!t]
	\centering
	\caption{Parameters for test cases.} 
	\renewcommand{\arraystretch}{1.0}
	\begin{threeparttable}
		\begin{tabular}{c|c|c|c|c|c|c}
			\toprule
			\hline
			Case & \tabincell{c}{$N$} & \tabincell{c}{$|\mathcal{G}|$} & \tabincell{c}{$|\mathcal{D}|$} &\tabincell{c}{$|\mathcal{E}|$} &\tabincell{c}{$N_{\scaleto{hid}{3pt}}$} &\tabincell{c}{Neurons per hidden layer}
			\\
			\hline
			\tabincell{c}{IEEE \\Case30} & 30 & 2 & 21 &41 & 3&32/16/8\\
			\hline
			\tabincell{c}{IEEE \\Case57} & 57 & 4 & 42 &80&3&32/16/8\\
			\hline
			\tabincell{c}{IEEE \\Case118} & 118 & 19 & 99 &186&3&128/64/32\\
			\hline
			\tabincell{c}{IEEE \\Case300} & 300 & 57 & 199 &411&3&256/128/64\\
			\hline
			\bottomrule
		\end{tabular}
		\begin{tablenotes}
			\footnotesize
			\item[*] A bus is considered a load bus if its default active power consumption is positive.
		\end{tablenotes}
	\end{threeparttable}
	\label{table1}
\end{table}

\begin{table*}[!t]
	\centering
	\caption{Performance comparison under typical operating conditions.}
	\renewcommand{\arraystretch}{1.0}
	\begin{threeparttable}
		\begin{tabular}{c|c|c|c|cc|c|cc|c}
			\toprule
			\hline
			\multirow{2}{*}{Test case} &
			\multirow{2}{*}{\tabincell{c}{\# Contingencies}} &
			\multirow{2}{*}{\tabincell{c}{\# Variables}} &
			\multirow{2}{*}{\tabincell{c}{Feasibility before\\ $\ell_1$-projection (\%)}} &
			\multicolumn{2}{c|}{\tabincell{c}{Average cost (\$/hr)}} &
			\multirow{2}{*}{\tabincell{c}{Optimality \\loss (\%)}} &
			\multicolumn{2}{c|}{\tabincell{c}{Running time (millisecond)}} &
			\multirow{2}{*}{\tabincell{c}{Speedup}} \\
			\cline{5-6} \cline{8-9}
			&&&&\textsf{DeepOPF}&Ref.&&\textsf{DeepOPF}&Ref.&\\
			\hline
			\tabincell{c}{IEEE Case30} &38& 1172&100 & 225.7  & 225.7 & $<$0.1 & 0.72 & 17 & $\times 24 $\\
			\hline
			\tabincell{c}{IEEE Case57} &79&4564& 100 & 9022.9 & 9021.6 & $<$0.1&0.76  & 102 & $\times 133$\\
			\hline
			\tabincell{c}{IEEE Case118}&177& 21023 & 100  &  29197.9&29149.0 & $<$0.2 & 2.48 &698 & $\times 281$ \\
			\hline
			\tabincell{c}{IEEE Case300}&318&95757&81.7 & 156601.8  & 156542.5 &  $<$0.1 &  81.4 &5766   & $\times 318 $ \\ 
			\hline			
			\bottomrule
		\end{tabular}
		\end{threeparttable}
	\label{table2}
\end{table*}

\section{Numerical Experiments}\label{sec:simulations}

\subsection{Experiment Setup}
\subsubsection{Simulation environment} The experiments are conducted in {CentOS 7.6} on the quad-core (i7-3770@3.40G Hz) CPU workstation and 16GB RAM. 
\subsubsection{Test case}
We consider four IEEE standard cases in the Power Grid Lib~\cite{babaeinejadsarookolaee2019power} (version 19.05): the IEEE Case- /30/57/118/300 test systems, representing small-scale, medium-scale, and large-scale power networks, respectively. Their illustrations are in~\cite{4077121,tpcwTrey6} and their parameters are shown in Table~\ref{table1}. For each case, we consider the typical operating conditions~\cite{babaeinejadsarookolaee2019power}, where the active power loads are within the normal region and the branch limits are not binding during both the pre-/post- contingency cases. Note the power flow balance constraints are active so the SC-DCOPF under the typical operating conditions are still a constrained optimization problem. We illustrate the detailed architecture of our DNN model for the IEEE Case30 in Fig.~\ref{fig4}.

\subsubsection{Data preparation}
In the training stage, the load data is sampled uniformly at random within $[90\%, 110\%]$ of the default value on each bus~\cite{babaeinejadsarookolaee2019power}. As the Power Grid Lib only has linear cost functions for generators, we use the cost functions from the test cases with same bus from MATPOWER~\cite{zimmerman2011matpower} (version 7.0) while all other parameters are taken from the Power Grid Lib cases. Then we obtain the solution of the SC-DCOPF problems by Gurobi~\cite{gurobi} (version 8.1.1). 
We sample 50,000 training data and 5,000 test data for each test case.

\subsubsection{The implementation of the DNN model}
We design the DNN model based on Pytorch platform and apply the stochastic gradient descent (SGD) method with momentum~\cite{qian1999momentum} to train the neural network. The epoch is set to 300 and the batch size is 64. {We set the weighting factors in the loss function in \eqref{equation13} to be $w_1=w_2=1$, based on empirical experience. The remaining parameters are shown in Table~\ref{table1}, including the number of hidden layers and the number of neurons per layer.} 

\subsubsection{Evaluation metrics}
We compare the performance of \textsf{DeepOPF} and the state-of-the-art Gurobi solver\footnote{{Gurobi implements the simplex algorithm for solving linear problems, which has a polynomial-time complexity with high probability~\cite{vaidya1989speeding} and a celebrated average-case running time performance}. The Gurobi solver by default uses multi-threading technique, which affects the computing time due to the threads' communication overhead. For fair comparison, we use the single-threading setting in our simulations.} using the following metrics, averaged over 5,000 test instances. The first is the percentage of the feasible solution obtained by both approaches. The second is the objective cost obtained by both approaches. The third is the running time, i.e., the average computation time for obtaining solutions for the 5,000 instances. The fourth is the speedup, i.e., the average of the running-time ratios of the Gurobi solver to \textsf{DeepOPF} for all the test instances. It captures the average gain in computation time, of using \textsf{DeepOPF} over the Gurobi solver. We note that the speedup is the average of ratios, and it is different from the ratio of the average running times between the Gurobi solver and \textsf{DeepOPF}.

\begin{table*}[!t]
	\centering
	\caption{Performance under the typical, lightly-congested, and heavily-congested settings.}
	\renewcommand{\arraystretch}{1.0}
	\resizebox{\textwidth}{!}{
		\begin{threeparttable}[b]
			\begin{tabular}{c|c|c|c|c|c|c|c|c|c|c}
				\toprule
				\hline
				\multirow{3}{*}{\tabincell{c}{Scheme}} &				
				\multirow{3}{*}{\tabincell{c}{Variants}}&
				\multicolumn{3}{c|}{\tabincell{c}{Typical}} &
				\multicolumn{3}{c|}{\tabincell{c}{Lightly-congested}} &
				\multicolumn{3}{c}{\tabincell{c}{Heavily-congested}} \\
				\cline{3-11}
                &&\tabincell{c}{Feasibility rate \\(\%)} & \tabincell{c}{Optimality  gap \\(\%)} & Speedup
				&\tabincell{c}{Feasibility rate \\(\%)} & \tabincell{c}{Optimality  gap \\(\%)} & Speedup
				&\tabincell{c}{Feasibility rate \\(\%)} & \tabincell{c}{Optimality  gap \\(\%)} & Speedup\\
				\hline           
				\multirow{3}{*}{\tabincell{c}{DNN}} &\tabincell{c}{with \\$\ell_1$-projection} &100 &$<$0.1 &338 &100 &$<$0.2 &56 &100 &$<$0.2 & $\times$16.4\\
                \cline{2-11}
                & \tabincell{c}{without \\$\ell_1$-projection}&100 &$<$0.1 &338 & 15.7&$<$0.2 &315 &0 &$<$0.2 & -- \\
                \hline
                \multirow{3}{*}{\tabincell{c}{KNN\\-50K }} &\tabincell{c}{with \\$\ell_1$-projection} &100 &$<$0.1 &0.5 &100 &$<$0.6 &0.7 &100 &$<$0.3 & $\times$1.5 \\
				\cline{2-11}                
                & \tabincell{c}{without \\$\ell_1$-projection}&100 &$<$0.1 &0.5 &0 &$<$0.9 &-- &0 &$<$0.3 & --\\
				\hline
				\bottomrule
			\end{tabular}
			\begin{tablenotes}
			\footnotesize
			\item[*] 
			`--' means the schemes fail to provide feasible solutions without post-processing thus do not associate with any speedup numbers. 
		\end{tablenotes}
		
	\end{threeparttable}}
	
	\label{table3.1}
\end{table*}

\subsection{Performance  under the Typical Operating Condition}
The simulation results for the test cases under the typical operating conditions are shown in Table~\ref{table2} and we have several observations. {First, as compared to the Gurobi solver, \textsf{DeepOPF} speeds up the computing time by up to two orders of magnitude. The speedup increases as the test cases get larger, suggesting that \textsf{DeepOPF} is more efficient for large-scale power networks.} Second, \textsf{DeepOPF} without involving the post-processing procedure always generates feasible solutions for IEEE Case30, IEEE Case57, and IEEE Case118, which justifies our design. We note that for IEEE Case300, \textsf{DeepOPF} achieves 81.7\% feasibility rate before the post-processing procedure and overall 318 average speedup. Further analysis shows the average speedup for the test instances with feasible solutions generated by DNN (thus without involving the post-processing procedure) is 385 with an average running time of 15ms. For the remaining 18.3\% test instances for which DNN generates infeasible solutions, it is due to the violation of 1 or 2 line capacity limit constraints. The $\ell_1$-projection based post-processing procedure is involved to obtain feasible solutions, and the average running time of \textsf{DeepOPF} with $\ell_1$-projection is 378ms. Overall, the average \textsf{DeepOPF} running time for all the IEEECase300 test instances is 81.4ms and the average speedup is 318. Third, the cost difference between the \textsf{DeepOPF} solution and the Gurobi solution is with less than 0.2\% optimality loss (on average). \rev{We show detailed statistics of the optimality loss and the speedup for the IEEE Case118, in Appendix~\ref{apx:cdf} of the supplementary materials, to further demonstrate the effectiveness of the \textsf{DeepOPF}. As compared to the optimal solution obtained by the Gurobi solver, \textsf{DeepOPF} achieves an average optimality loss less than 0.2\% with the maximum around 1.2\%. Meanwhile, \textsf{DeepOPF} achieves an average speedup of $\times$281 with the maximum around $\times$320. More details can be found in Appendix~\ref{apx:cdf} of the supplementary materials.}

\subsection{Performance with High-Variation Load and under Congested Settings}
\label{ssec:high-var-load-and-congested}
To stress-test \textsf{DeepOPF}, we enlarge the sampling range of the load on each bus and carry out simulations on IEEE Case118 under the typical, lightly-congested, and heavily-congested settings, by using and adjusting the typical and congested configurations of IEEE Case118 provided by the Power Grid Lib. 
\rev{For each setting, we sample 50,000 data in the load region for training and prepare another 5,000 for testing. For comparison, we evaluate the performance of the KNN scheme with $K=50$, also using the same 50,000 sample data and 5,000 test data under each setting for fair comparison. We denote the scheme as KNN-50K. Its output is calculated as the average of the generation profiles (except the slack bus) of the $K$ nearest neighbors of the input load in the training data set. Then the slack bus generation is computed to ensure the loads are satisfied. The phase angles on each bus can be uniquely determined by solving the power flow equations in~\eqref{equation5}. The feasibility of the power flow on each line is evaluated by~\eqref{equation12}.}

The results are reported in Table~\ref{table3.1}. For the typical setting, the load variation region is set as $[50\%, 150\%]$ of the default load and the line and generation limits are set according to the typical setting provided by the Power Grid Lib. Under such a setting, we observe none of the line constraints are binding in the test data set. As seen from the Table~\ref{table3.1},  \textsf{DeepOPF} achieves a 100\% feasibility rate, decent speedup, and 0.1\% optimality loss. This implies \textsf{DeepOPF} works well on the high-variation load under the typical operating setting. \rev{Meanwhile, it achieves similar optimality gap performance as the KNN scheme, but a better speedup performance.}
    
For the lightly-congested setting, the load variation region is set as $[50\%, 150\%]$ of the default load and the line and generation limits are set according to the congested setting in the Power Grid Lib. Under the setting, 85\% of test cases have at least one line constraint binding. Under this setting, \textsf{DeepOPF} with the post-processing procedure generates feasible solutions with a 0.2\% optimality loss and a $ \times $56 speedup as compared to the Gurobi solver. For the 85\% test cases with at least one line constraint binding, \textsf{DeepOPF} with the post-processing procedure obtains solutions with less than 0.2\% optimality loss and a $\times$8 speedup. \rev{As compared to the KNN scheme, \textsf{DeepOPF} achieves better speedup and optimality gap performance. We note that the absolute load range under this lightly-congested setting is larger than the other two settings. Consequently, the samples are sparser, resulting in the worst optimality gap performance of KNN-50K, even worse than that under the heavily-congested setting.}
    
For the heavily-congested setting, we set the load variation region to be within $[150\%, 160\%]$ and adjust the line flow limits under the largest load input. With the adjustment, all the 50,000 training and 5,000 test instances have about 20\% line constraints binding. We note that the adjustment requires tuning the line limits to have as many lines constraints binding but without introducing post-contingency infeasibility. As we see from the simulation results,  \textsf{DeepOPF} without post-processing fails to generate feasible solutions. In contrast, \textsf{DeepOPF} with post-processing generates 100\% feasible solutions, with \rev{less than $0.2\%$} optimality loss and a \rev{$\times$16} speedup as compared to the Gurobi solver. \rev{Under the heavily-congested setting, both KNN and \textsf{DeepOPF} scheme fail to generate feasible solutions for the test instances. After applying the $l_1$-projection based post-processing procedure, both schemes obtain feasible solutions and \textsf{DeepOPF} achieves better overall speedup performance and optimality loss performance.}

Overall, under both lightly-congested and heavily-congested settings, our simulation results show that the post-processing step is more efficient than solving the original problem directly. \rev{The results also echo the general understanding that KNN incurs less training-time complexity but higher running-time complexity than neural-network based approaches.}

\begin{figure} [!ht]
  \centering
  \subfigure[]{
    \includegraphics[width = 0.45\linewidth]{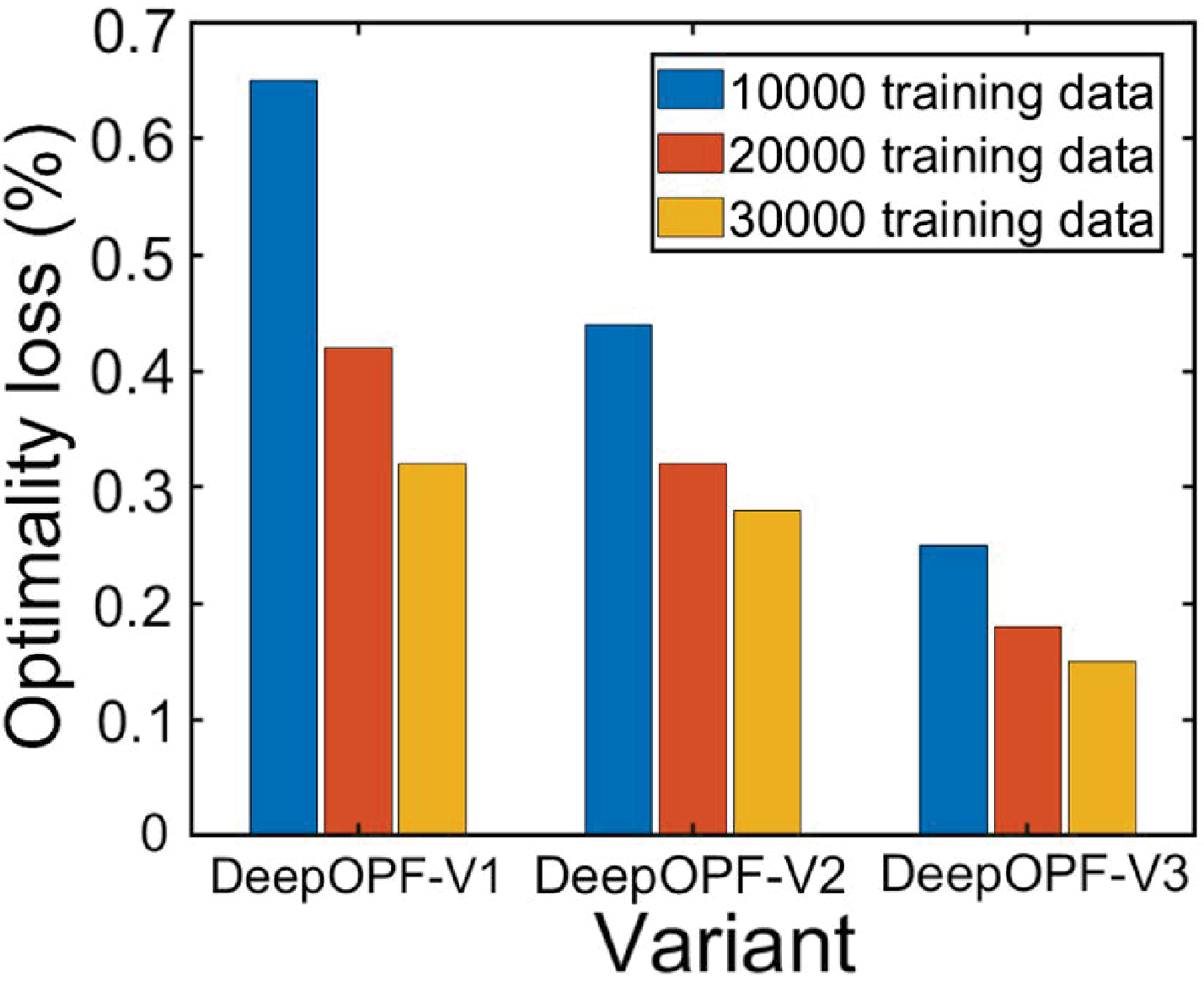}\label{figure4_a}
  }
  \subfigure[]{
    \includegraphics[width = 0.45\linewidth]{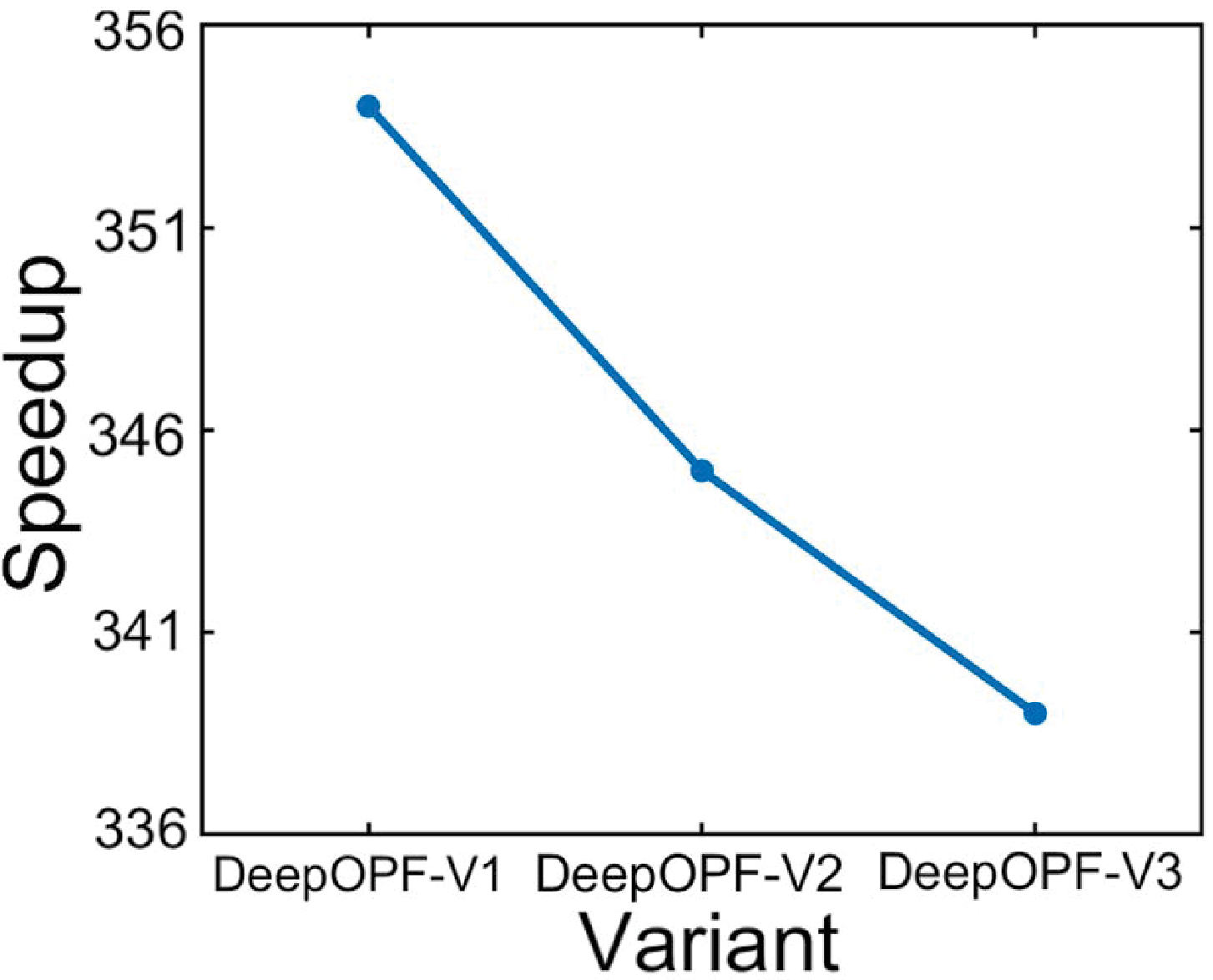}\label{figure4_b}
  }
  \caption{Performance under different neural network and training data sizes for IEEE Case118 under typical operating conditions.}\label{figure5}
\end{figure}

\subsection{Performance with different DNN Scales and Training data Sizes}
When applying DNN approaches, it is of interest to evaluate the influence of the DNN's size and the amount of training data on the performance. In addition to the corresponding performance analysis w.r.t. the DNN's size in Sec.~\ref{sec:OPF.theory_analysis2}, we carry out experiments to compare the optimality loss and speedup of \textsf{DeepOPF} with different neural network size and training data size for IEEE case118 under the typical operation condition. Three DNN models of different scales are used for comparison: 
\begin{itemize}
\item \textsf{DeepOPF-V1}: A simple neural network with one hidden layer; the number of neurons is 16.
\item \textsf{DeepOPF-V2}:  A simple neural network with two hidden layers; the numbers of neurons per layer are 32 and 16, respectively.
\item \textsf{DeepOPF-V3}: A simple neural network with three hidden layers; the numbers of neurons per layer are 64, 32, and 16, respectively.
\end{itemize}
The training data size varies from 10,000 to 30,000. The results are shown in Fig.~\ref{figure4_a} and Fig.~\ref{figure4_b}. It is observed that larger training data size contributes to smaller optimality loss. Furthermore, we observe that when the depth and the size of the neural network increase, \textsf{DeepOPF} achieves better performance on optimality loss but less speedup. The above results correspond to our theoretical analysis on computational complexity and prediction accuracy in Sec.~\ref{ssec:comp.complexity} and Sec.~\ref{sub_approximation_error}, i.e., larger DNN size tends to have better prediction accuracy (smaller optimality loss) but also higher computational complexity. Having said so, the over-fitting issue may appear in practice if we keep increasing the depth and size. Thus, for different power networks (as IEEE test cases), the DNN model can be determined by educated guesses and iterative tuning, which is also by far the common practice in generic DNN approaches in various engineering domains.

\begin{table}[!t]
\caption{Performance comparisons of different combinations of weights in the loss function.}
	\renewcommand{\arraystretch}{1.1}
	\centering
	\begin{tabular}{c|c|c|c|c}
		\toprule
		\hline
        \multicolumn{2}{c|}{Weight setting} & \tabincell{c}{Feasibility \\rate (\%)} & \tabincell{c}{Optimality  \\loss (\%)} & Speedup \\
        \hline
        \multirow{3}{*}{\tabincell{c}{$w_1 = 1$, \\ $w_2 = 1$}} &\tabincell{c}{with \\$\ell_1$-projection} &100 &$<$0.2 & $\times$56 \\
        \cline{2-5}
        & \tabincell{c}{without \\$\ell_1$-projection}& 15.7& $<$0.2&$\times$315\\
        \hline
        \multirow{3}{*}{\tabincell{c}{$w_1 = 1$, \\ $w_2 = 10$ }} &\tabincell{c}{with \\$\ell_1$-projection} &100 & $<$0.3& $\times$83\\
        \cline{2-5}
        & \tabincell{c}{without \\$\ell_1$-projection}&23.8& $<$0.3&$\times$324\\
        \hline
        \multirow{3}{*}{\tabincell{c}{$w_1 = 10$, \\ $w_2 = 1$}} &\tabincell{c}{with \\$\ell_1$-projection} &100 &$<$0.1 & $\times$53 \\
        \cline{2-5}
        & \tabincell{c}{without \\$\ell_1$-projection}&14.5&$<$0.1 & $\times$324\\
		\hline
		\bottomrule		
		\end{tabular}
	\label{table4}
\end{table}

\subsection{Performance with different Weighting Factors in Loss Function}
As shown Sec.~\ref{ssec:DNN.and.loss.function}, there are two weighting factors $w_1$ and $w_2$ in the loss function to balance between the training loss and the penalty of violating the inequality constraints. We carry out comparative experiments to evaluate the influence of the two hyper-parameters on the performance. More specifically, we use IEEE Case118 with 50\% sampling range for testing, where the penalty is more likely to take effect as several transmission lines are binding. Three variants of the weighting factors in the loss function and the corresponding results are shown in Table~\ref{table4}.
As seen, larger value of $w_2$ enhances the feasibility rate (before $\ell_1$-projection) and the speedup as the post-processing step is involved in fewer test instances. In practice, the weight factors can be determined by educated guesses and iteratively adjusted to balance the influence of the two term in the loss function.
\section{Conclusion}\label{sec:conclusion}
\rev{We develop \textsf{DeepOPF} for solving  SC-DCOPF problems. Given a power network, \textsf{DeepOPF} employs a DNN to learn a high-dimensional mapping between the load inputs and the dispatch decisions. With the learned mapping, it first obtains the generations from the load inputs and then directly computes the phase angels from the generations and loads. We also develop an $\ell_1$-projection based post-processing procedure  to ensure the feasibility of the obtained solution. We analyze the approximation capability and computational complexity of \textsf{DeepOPF}. Simulation results show that \textsf{DeepOPF} generates feasible solutions with less than \rev{0.2\%} optimality loss. As compared to the Gurobi solver, \textsf{DeepOPF} speeds up the computation time by up to two orders of magnitude under the typical operating condition and by up to one order of magnitude under the congested condition. The approach may be computationally expensive in constructing and training the DNN model, which can be minor if amortized over many problem instances (e.g., one per every 5 minutes) that can be solved using the same model. Future directions include evaluating \textsf{DeepOPF} for national-scale power transmission networks and extending it to the AC-OPF setting~\cite{pan2020deepopf}.}

\bibliographystyle{IEEEtran}
\bibliography{IEEEabrv,ref}

%

\begin{IEEEbiography}[{\includegraphics[width=1in,height=1.25in,clip,keepaspectratio]{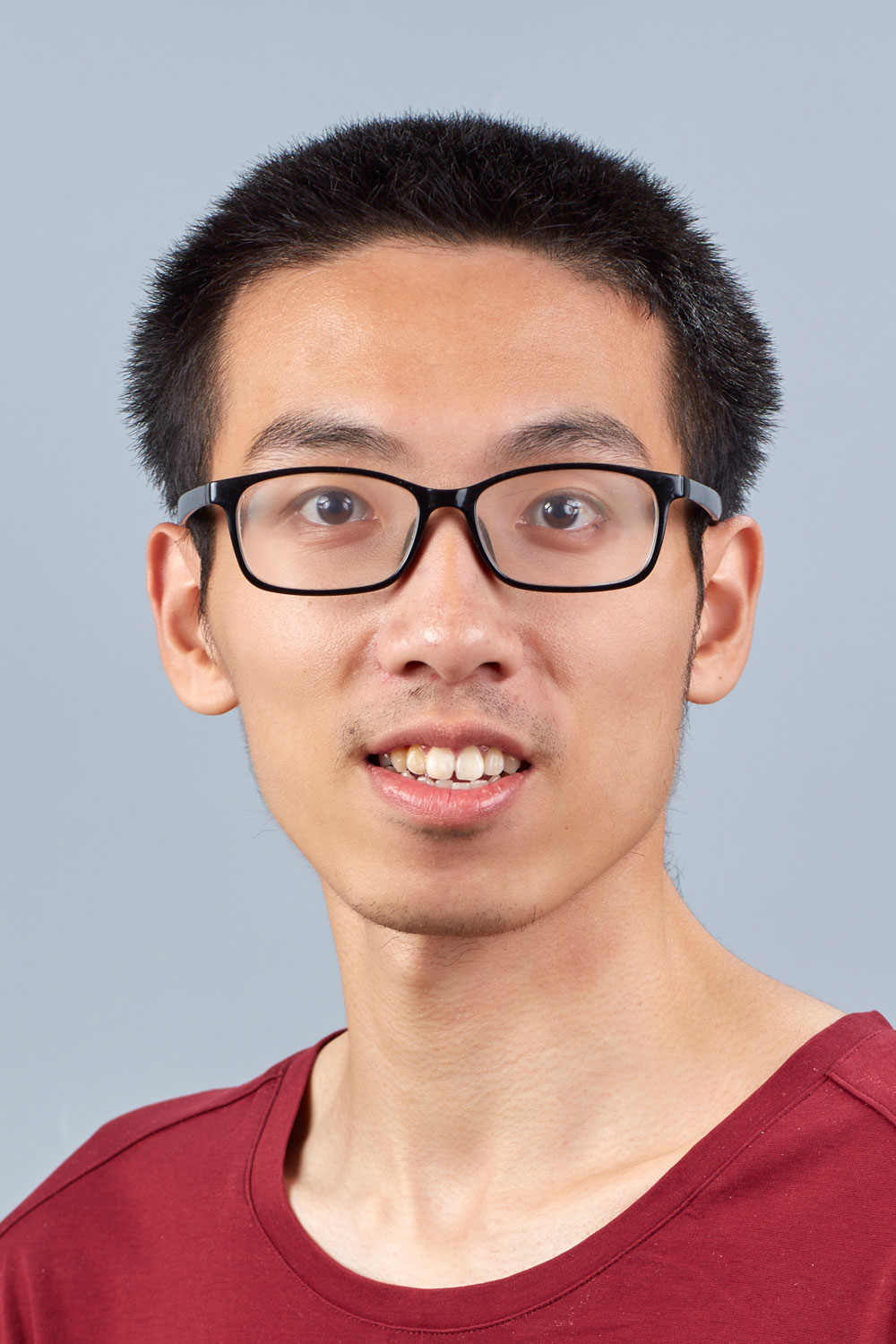}}]
{Xiang Pan} (S'18) received the M.S.\ degree in computer science from Wuhan University, Wuhan, China, in 2018. He is currently pursuing the Ph.D. degree with the Department of Information Engineering, The Chinese University of Hong Kong. His research interest includes machine learning and its application in power systems.
\end{IEEEbiography}

\vskip -1\baselineskip plus -1fil

\begin{IEEEbiography}[{\includegraphics[width=1in,height=1.25in,clip,keepaspectratio]{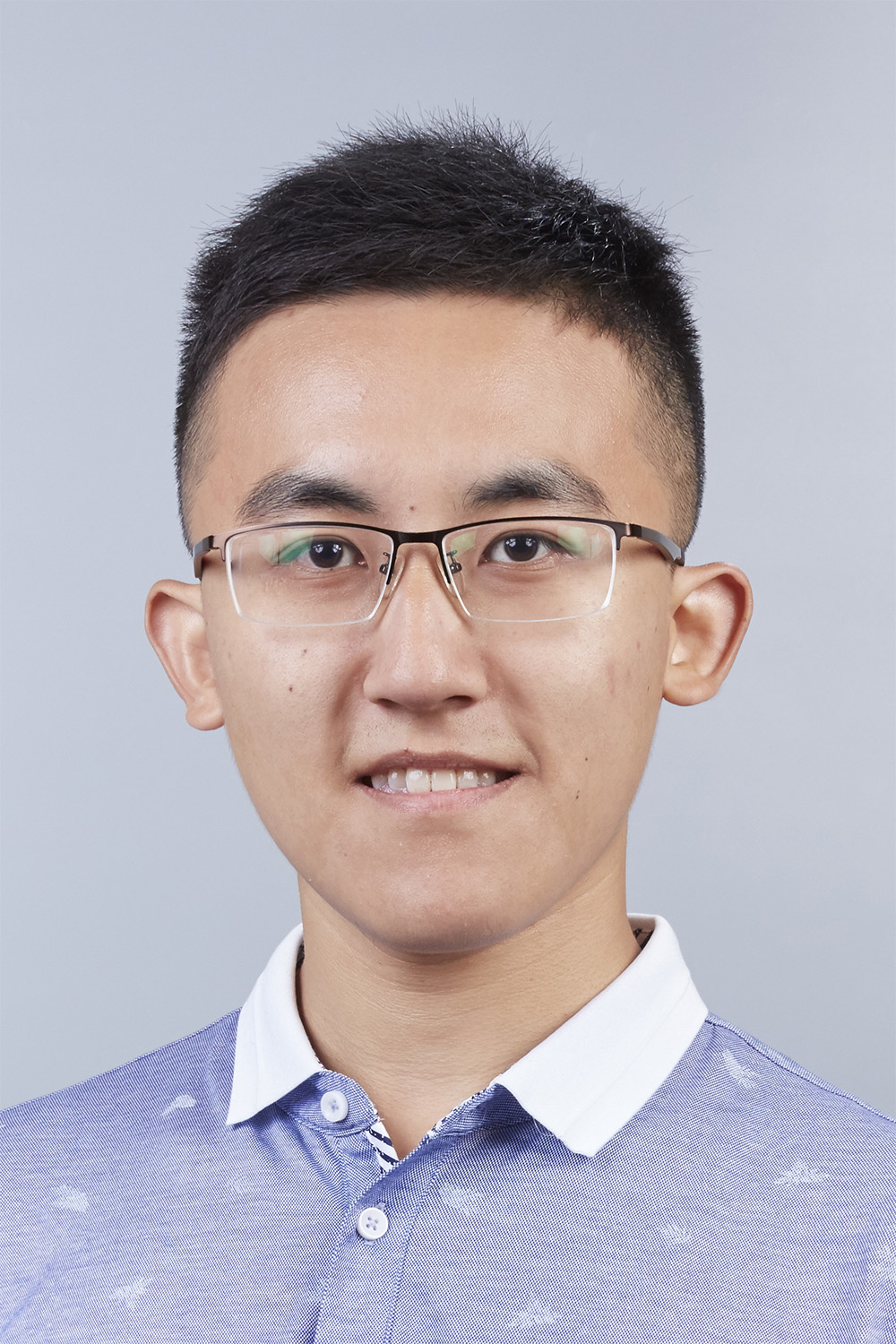}}]
{Tianyu Zhao} received the B.Eng. degree from the School of Energy and Power Engineering and Minor's Diploma in Business Administration from the School of Management, Xi'an Jiaotong University, Xi'an, China, in 2017. He is currently pursuing the Ph.D. degree with the Department of Information Engineering, The Chinese University of Hong Kong. His research interests include electricity market, microgrid operation, and machine learning applications in power systems.
\end{IEEEbiography}

\vskip -1\baselineskip plus -1fil

\begin{IEEEbiography}[{\includegraphics[width=1in,height=1.25in,clip,keepaspectratio]{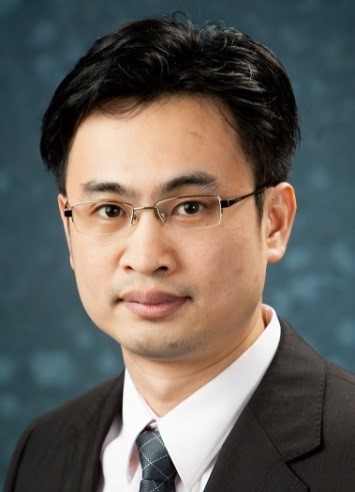}}]
{Minghua Chen} (S'04-M'06-SM'13) received his B.Eng. and M.S. degrees from the Dept. of Electronic Engineering at Tsinghua University. He received his Ph.D.\ degree from the Dept.\ of Electrical Engineering and Computer Sciences, University of California Berkeley. He is currently a Professor in the School of Data Science, City University of Hong Kong. He received the Eli Jury award from UC Berkeley in 2007 (presented to a graduate student or recent alumnus for outstanding achievement in the area of Systems, Communications, Control, or Signal Processing) and The Chinese University of Hong Kong Young Researcher Award in 2013. He also received several best paper awards, including the IEEE ICME Best Paper Award in 2009, the IEEE Transactions on Multimedia Prize Paper Award in 2009, and the ACM Multimedia Best Paper Award in 2012. He also co-authors several papers that are Best Paper Award Runner-up/Finalist/Candidate for ACM MobiHoc in 2014 and ACM e-Energy in 2015, 2016, 2018, and 2019. Prof. Chen serves as TPC Co-Chair, General Chair, and Steering Committee Chair of ACM e-Energy in 2016, 2017, and 2018-present, respectively. He also serves as Associate Editor of IEEE/ACM Transactions on Networking in 2014-2018. He receives the ACM Recognition of Service Award in 2017 for the service contribution to the research community. He is currently serving as TPC Co-Chair for ACM MobiHoc 2020. His current research interests include online optimization and algorithms, energy systems (e.g., smart power grids and energy-efficient data centers), intelligent transportation systems, distributed optimization, delay-constrained network communication, and capitalizing the benefit of data-driven prediction in algorithm/system design.
\end{IEEEbiography}

\vskip -1\baselineskip plus -1fil

\begin{IEEEbiography}[{\includegraphics[width=1in,height=1.25in,clip,keepaspectratio]{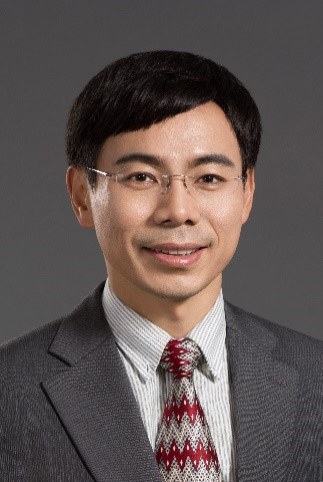}}]
{Shengyu Zhang} obtained his bachelor degree in mathematics, Fudan University in 1999, master in computer science, Tsinghua University in 2002, under the supervision of Mingsheng Ying, and Ph.D. in computer science, Princeton University in 2006, under the supervision of Prof. Andrew Yao. He then worked in NEC Laboratories America as a summer intern, and in California Institute of Technology for a two-year postdoc, hosted by Prof. John Preskill, Prof. Leonard Schulman and Prof. Alexei Kitaev. He joined The Chinese University of Hong Kong as an assistant professor in 2008, and became an associate professor in 2014. In January 2018, he joined Tencent as a Distinguished Scientist, in charge of Tencent Quantum Laboratory.

Shengyu Zhang's research interest lies in quantum computing, algorithm designing, and foundation of artificial intelligence. He is an editor of Theoretical Computer Science, and of International Journal of Quantum Information. He published numerous papers and served as PC member in leading conferences in theoretical computer science, quantum computing, and artificial intelligence. 
\end{IEEEbiography}





\clearpage

\section*{\large \textbf{Supplementary Materials}}

\begin{appendices}

\section{Proof of Lemma~\ref{lem1}} \label{lem1_proof}
\begin{proof}
We now show the considered piece-wise linear {one-dimensional} output function $f^*(\cdot)$ is Lipschitz-continuous in the input domain $\mathcal{S}$, which can be partitioned into $r$ different convex polyhedral regions, ${R}_i,i=1,...,r$. The mapping $f^{*}\left( \cdot \right)$ is  piece-wise linear and  can be defined as follows: 
 \begin{equation*}
     f^{*}\left( x \right)=
     \begin{cases}
     a_1 x + b_1, \ \textrm{if} \ x \in {R}_1;\\
     a_2 x + b_2, \ \textrm{if} \ x \in {R}_2;\\
     \cdots \\
     a_r x + b_r, \ \textrm{if} \ x \in {R}_r;\\
     \end{cases}
 \end{equation*}
where $x \in \mathbf{R}^{n \times 1}, a_i \in \mathbf{R}^{1 \times n}, i=1,...,r $ and $b_i \in \mathbf{R}^{1}, i=1,...,r $. Then, we can have:
 \begin{equation}
 \begin{aligned}
 \left| f^{*}\left( x_1 \right) -f^{*}\left( x_2 \right) \right| \le \lVert a_i \rVert \cdot \lVert x_1-x_2 \rVert, \;\;\forall x_1,x_2 \in \mathcal{S}. \nonumber
 \end{aligned}
 \end{equation}
 Thus, let  $ \varLambda = \max\left\{ \lVert a_i \rVert, \dots, \lVert a_r \rVert \right\}$. We have
 \begin{equation}
 \begin{aligned}
 \left| f^{*}\left( x_1 \right) -f^{*}\left( x_2 \right)  \right| \le \varLambda \cdot \lVert x_1-x_2 \rVert, \;\; \forall x_1,x_2 \in \mathcal{S}.\nonumber
 \end{aligned}
 \end{equation}
Therefore, $f^{*}\left( \cdot \right)$ is Lipschitz-continuous.
\end{proof}

\section{Proof of Lemma~\ref{lem3}} \label{apx:proof_of_lemma_3}
Before we proceed, we present a result on the approximation error between two scalar function classes.
\begin{lemma}\label{lem3}
Let $\mathcal{H}$ be the class of two-segment piece-wise linear functions with a Lipschitz constant $\varLambda>0$, over an interval $\left[-\mu,\mu \right]$ ($\mu>0$). Let $\mathcal{K}$ be the class of all linear scalar functions over $\left[-\mu,\mu \right]$. Then, the following holds,
\begin{equation}\label{equation17}
\underset{h\in \mathcal{H}}{\max}\ \underset{g\in \mathcal{K}}{\min} \ \underset{x\in \left[-\mu ,\mu \right]}{\max}\left|h\left( x \right) -g\left( x \right) \right|\geq \varLambda \cdot \frac{\mu}{2}.
\end{equation}
\end{lemma}
Essentially, the lemma gives a lower bound to the worst-case error of using a linear function to approximate a two-segment piece-wise linear function.  
\begin{proof}\label{proof5}
We can derive the lower bound to the worst-case $L_{\infty}$-based approximation error as follows. Suppose we want to find a function $g \left( \cdot \right)$ belongs to the linear scalar function class $\mathcal{K}$ to approximate the function $h$ belongs to the two-segment piece-wise linear function class $\mathcal{H}$ with a Lipschitz constant $\varLambda > 0$, over an interval $\left[-\mu,\mu \right]$ ($\mu>0$). An illustration is shown in Fig.~\ref{fig_appro}. Let ${g\left( x \right)}=a\cdot x+b$, for ${x\in \left[ -\mu, \mu \right]}$. Let $\hat{h} \in \mathcal{H}$ be the following:
\begin{equation}
    \hat{h} \left(x \right)=
    \begin{cases}
    \varLambda (x+\mu), \ \textrm{if} \ x \in \left[-\mu,0 \right];\\
    -\varLambda (x-\mu), \ \textrm{if} \ x \in \left[0, \mu \right];\\
    \end{cases}
\end{equation}

Then, we can obtain the lower bound for the $L_{\infty}$-based approximation error of $\hat{h} \left( \cdot \right)$ and $g\left( \cdot \right)$ by the classification discussion on the intercept $b$.
\begin{itemize}
\item If $b \le \frac{\varLambda \mu}{2}$. Under this case, we can get: 
\begin{eqnarray*}
\underset{x\in \left[-\mu ,\mu \right]}{\max}\left| \hat{h}\left( x \right) -g\left( x \right) \right|&\geq& \left| \hat{h}\left( 0 \right) -g\left( 0 \right) \right|\\
&\geq& \varLambda \cdot \frac{\mu}{2n}\\
&=&\left| \varLambda \mu -b \right|\\
&\geq& \varLambda \cdot \frac{\mu}{2}.
\end{eqnarray*}

\item Otherwise $\frac{\varLambda \mu}{2}< b $. If $a>0$, under this case we can have:
\begin{eqnarray*}
\underset{x\in \left[-\mu ,\mu \right]}{\max}\left| \hat{h}\left( x \right) -g\left( x \right) \right|&\geq& \left| \hat{h}\left( \mu \right) -g\left( \mu \right) \right|\\
&\geq& \left( \varLambda +a \right) \cdot \frac{\mu}{2}\\
&\geq& \varLambda \cdot \frac{\mu}{2}.
\end{eqnarray*}
Otherwise $a\leq 0$, we can consider the point $x=-\mu$ and obtain the same result.
\end{itemize}
Thus overall, we observe
\begin{equation}
\underset{g\in \mathcal{K}}{\min}\ \underset{x\in \left[-\mu ,\mu \right]}{\max}\left| \hat{h}\left( x \right) -g\left( x \right) \right|= \varLambda \cdot \frac{\mu}{2}.\nonumber
\end{equation}
For the worst-case $L_{\infty}$-based approximation error, we have
\begin{align}
    & \underset{h\in \mathcal{H}}{\max}\ \underset{g\in \mathcal{K}}{\min}\ \underset{x\in \left[-\mu ,\mu \right]}{\max}\left| h\left( x \right) -g\left( x \right) \right| \nonumber \\
    \geq& \underset{g\in \mathcal{K}}{\min}\ \underset{x\in \left[-\mu ,\mu \right]}{\max}\left| \hat{h}\left( x \right) -g\left( x \right) \right| \nonumber \\
    \geq& \varLambda \cdot \frac{\mu}{2}. \nonumber
\end{align}

\end{proof}


\begin{figure}[t]
	\centering
	\includegraphics[width = 0.45\textwidth]{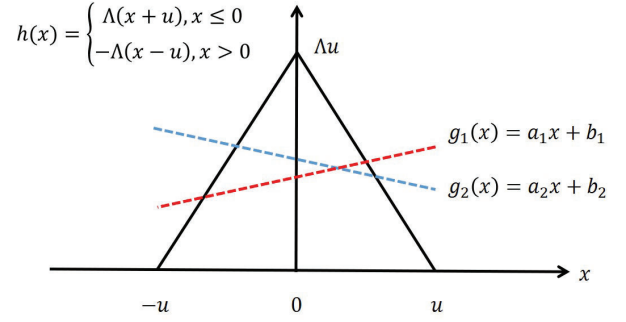}
	\caption{Illustration of approximating a two-segment piece-wise Lipschitz-continuous function $h(\cdot)$ by a linear function $g(\cdot)$.} 
	\label{fig_appro}
\end{figure}

\section{Proof of Theorem~\ref{thm:worst-cast-approximation-error}} \label{apx:proof_of_theory_1}
\begin{proof}

Suppose $\mathcal{K}$ is the family of piece-wise linear functions generated by a neural network with depth $N_{\scaleto{hid}{3pt}}$ and maximum number of neurons per layer $M$, on the load input domain $\mathcal{S}$ with the diameter $d$. The maximum number of segments any functions in $\mathcal{K}$ can have is defined as $n$. 
Let $\mathcal{H}$ be the class of all possible $f^*(\cdot)$ with a Lipschitz constant $\varLambda>0$. Let $[a_i, a_{i+1}]$, $0\leq i \leq 2n-1$, be $2n$ intervals with equal length portioning the diameter of input domain. Define $\hat{f} \in \mathcal{H}$ as follows:
\begin{equation*}
    \hat{f} \left(x \right)=
    \begin{cases}
    \varLambda (x-a_i), \ \textrm{if} \ x \in \left[a_i,a_{i+1} \right], i = 0, 2, \ldots, 2n-2;\\
    -\varLambda (x-a_{i+2}), \ \textrm{if} \ x \in\left[a_{i+1},a_{i+2} \right], i = 0, 2, \ldots, 2n-2.\\
    \end{cases}
\end{equation*}
Consider any $f \in \mathcal{K}$, since $f$ is piece-wise linear with at most $n$ segments over the input domain, it must be linear over one of the following $n$ segments $\left[a_i,a_{i+2} \right], i=0,2,...,2n-2$. Over that particular segment, we apply Lemma \ref{lem3} to bound the approximation error as in (\ref{equation17}).
Overall, we have
\begin{eqnarray}
&&\underset{f\in \mathcal{K}}{\min}\ \underset{x\in \mathcal{S}}{\max}\left| \hat{f}\left( x \right) -f\left( x \right) \right| \geq \varLambda \cdot \frac{d}{4n}.
\end{eqnarray}
Since the above inequality holds for a particular choice of $\hat{f}\in \mathcal{H}$, we must have
\begin{eqnarray}\label{equation19}
&&\underset{f^{*}\in \mathcal{H}}{\max}\,\, \underset{f\in \mathcal{K}}{\min}\ \underset{x\in \mathcal{S}}{\max}\left| f^{*}\left( x \right) -f\left( x \right) \right| \geq \varLambda \cdot \frac{d}{4n}.
\end{eqnarray}

Meanwhile, we use the result in \cite{telgarsky2016benefits}, of which the following is an immediate corollary.
\begin{corollary}
The maximum number of linear segments generated from the family of ReLU neural networks with depth (the number of hidden layers) $l$ and maximal width (neurons on the hidden layer) $m$ is $\left( 2m \right) ^l$.
\end{corollary}

By the above corollary, we have $n\leq \left(2M\right)^{N_{\scaleto{hid}{3pt}}}$. Plugging the relationship into (\ref{equation19}), we have
\begin{equation}
\underset{f^*\in \mathcal{H}}{\max}\ \underset{f\in \mathcal{K}}{\min}\ \underset{x\in \mathcal{S}}{\max}\left| f^*\left( x \right) -f\left( x \right) \right|\geq \varLambda \cdot \frac{d}{4\cdot(2M)^{N_{\scaleto{hid}{3pt}}}}.
\end{equation}
\end{proof}

\section{Proof of Corollary~\ref{col1} } \label{apx:proof_of_col1}
\begin{proof}
We next will show how to derive the Corollary \ref{col1}. Suppose $\epsilon$ is defined as the upper bound for the worst-case approximation error, that is:
\begin{equation}
\underset{f^*\in \mathcal{H}}{\max}\,\,\underset{f \in \mathcal{G}}{\min}\ \underset{x\in \mathcal{D}}{\max}\left| f^*\left(x\right) -f\left( x \right) \right|\le \epsilon
\end{equation}
Then, we can derive the following inequality based on the above definition and Theorem \ref{thm:worst-cast-approximation-error}:
\begin{equation}
\varLambda \cdot \frac{d}{4\cdot(2M)^{N_{\scaleto{hid}{3pt}}}} \le \epsilon,
\end{equation}
After some transformations, we can obtain the following necessary condition related to the DNN's scale on the Corollary \ref{col1}, which can guarantee that the designed DNN's ever possible to approximate the most difficult load-to-generation mapping with a Lipschitz constant $\varLambda$, up to an error of $\epsilon>0$:
\begin{equation}
(2M)^{N_{\scaleto{hid}{3pt}}}\geq \varLambda \cdot \frac{d}{4\cdot \epsilon}.
\end{equation}
\end{proof}

\section{Computational complexity of \textsf{DeepOPF} for Predicting the Generations} \label{apx:Complexity_DNN_prediction}
Recall that the number of bus and the number of contingencies are $N$ and $C$, respectively. The input and the output of the DNN model have $K_{\scaleto{in}{3pt}}$ and $K_{\scaleto{out}{3pt}}$ dimensions, and the DNN model has $N_{\scaleto{hid}{3pt}}$ hidden layers and each hidden layer has at most $M$ neurons. Specifically, in our setting, $K_{\scaleto{in}{3pt}}$ equals to the number of buses with load and $K_{\scaleto{out}{3pt}}$ equals to the number of generators. Therefore, the input and output dimensions are of the same order of $N$. From empirical experience, we set $M$ to be on the same order of $N$ and set $N_{\scaleto{hid}{3pt}}$ to be a constant. Once we finish training the DNN model, the complexity of generating solutions by using \textsf{DeepOPF} is characterized in the following proposition.
\begin{prop} \label{prop:DeepOPF.complexity}
The computational complexity (measured as the number of arithmetic operations) to generate the generations to the SC-DCOPF problem by using \textsf{DeepOPF} is 
\begin{equation}
    T=K_{\scaleto{in}{3pt}} K_1+ \sum^{N_{\scaleto{hid}{3pt}}-1}_{i=1}K_iK_{i+1} + K_{\scaleto{out}{3pt}} N_{\scaleto{hid}{3pt}},
    \label{complexity1}
    \end{equation}
which is $\mathcal{O}\left({N_{\scaleto{hid}{3pt}}} M^2\right)$.
\end{prop}
Note that $N_{\scaleto{hid}{3pt}}$ is set to 3 and $M$ is set to be $\mathcal{O}\left(N\right)$. {The complexity of \textsf{DeepOPF} for predicting the generations is $\mathcal{O}\left(N^2\right)$, 
smaller than that of the interior point method.} 

\begin{proof}
We next will show how to derive the computational complexity of using the DNN model to obtain the generation output from the given input. Recall that the input and the output of the DNN model in \textsf{DeepOPF} are $K_{\scaleto{in}{3pt}}$ and $K_{\scaleto{out}{3pt}}$ dimensions, respectively, and the DNN model has $N_{\scaleto{hid}{3pt}}$ hidden layers and each hidden layer has $K_{\mbox{i}}$ neurons, for $i=1,...,N_{\scaleto{hid}{3pt}}$. The maximal neurons on the hidden layers is $M$ neurons. For each neuron in the DNN model, we can regard the computation complexity on each neuron (measured by basic arithmetic operation) as  $\mathcal{O} \left(1\right)$. As we apply the fully-connected architecture, the output of each neuron is calculated by taking a weighted sum of the output from the neurons on the previous hidden layer and passing through a activation function.

Thus, the computational complexity (measured as the number of arithmetic operations) to generate the output from the input by a DNN model consists of the following three parts:
\begin{itemize}
\item Complexity of computation from the input to the first hidden layer. As each neuron on the first hidden layer will take the input data, thus the corresponding complexity  is $\mathcal{O} \left(K_{\scaleto{in}{3pt}}K_1\right)$.

\item Complexity of computation between the consecutive hidden layers. Since each neuron on the current hidden layer will take the output from each neuron on the previous hidden layer as the input data. Thus, thus the corresponding complexity is $\mathcal{O} \left(\sum_{i=1}^{N_{\scaleto{hid}{3pt}}-1}{K_iK_{i+1}}\right)$.

\item Complexity of computation from the last hidden layer to the output. As the output of each neuron on the last hidden layer is used to calculated the output, the corresponding complexity is $\mathcal{O} \left({N_{\scaleto{hid}{3pt}}}K_{\scaleto{out}{3pt}}\right)$.
\end{itemize}

The Sigmoid function is applied to each element of the output in order to guarantee that the elements of the final output is within $(0, 1)$. The Sigmoid function takes the form of
$$\displaystyle S(x)={\frac {1}{1+e^{-x}}}={\frac {e^{x}}{e^{x}+1}}.$$
and computing a Sigmoid function involves one addition operation, one division operation, and one exponentiation operation. The exponentiation operation is essentially a combination of $n$-th power operation and $m$-th root operation, where $n$ and $m$ are some integers depending on the output element $x$. That is, $x=\frac{n}{m}$. Previous works show that the computational complexity of $n$-th multiplication operations and $m$-th root operations is $\mathcal{O}(\text{log} n\cdot\text{log} m)$~\cite{chen1989fast,gordon1998survey}. Therefore, a Sigmoid function requires $\mathcal{O}(\text{log} n\cdot\text{log} m)$ operations. In practice, both $n$ and $m$ are bounded by some constant integer $M$ in the actual computation process, and therefore the computational complexity for the Sigmoid function is $(\text{log} M)^2$, which is a constant too. In our \textsf{DeepOPF} solution, the output layer of DNN has $K_{\scaleto{out}{3pt}}$ neurons with Sigmoid function, the corresponding computational complexity (the number of arithmetic operations) for the output layer is $\mathcal{O}(K_{\scaleto{out}{3pt}})$.

Hence, the overall complexity of the calculation by a DNN model is:
\begin{eqnarray*}
T&=&\mathcal{O}\left( N_{\scaleto{in}{3pt}}K_1+N_{\scaleto{hid}{3pt}}M^{2}+N_{\scaleto{hid}{3pt}}K_{\scaleto{out}{3pt}} \right)\\
&=&\mathcal{O}\left( N_{\scaleto{hid}{3pt}}M^{2} \right).\\
\end{eqnarray*}
\end{proof}

\begin{figure} [tbh]
  \centering
  \subfigure[]{
    \includegraphics[width = 0.46\linewidth]{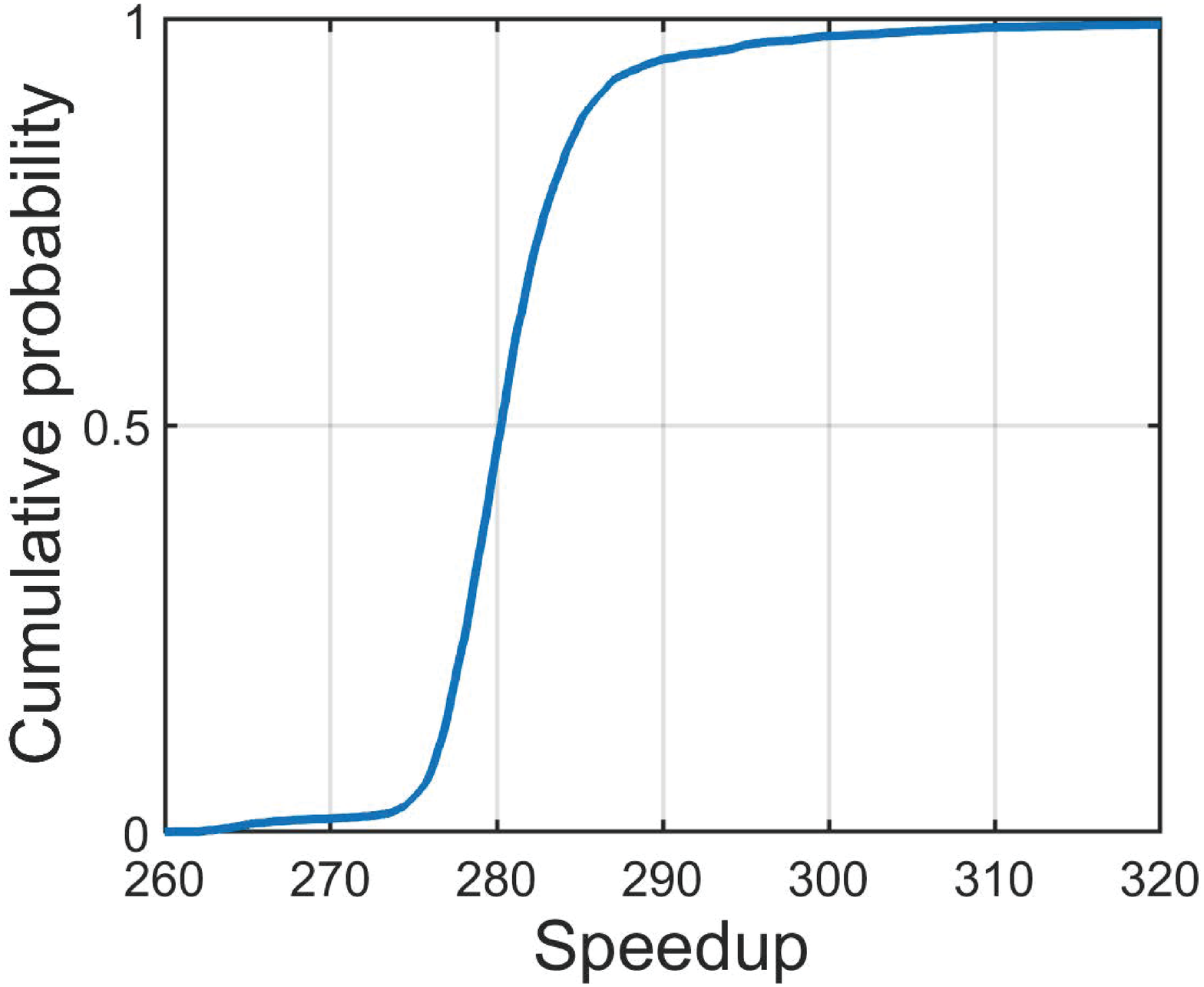}\label{figure6_a}
  }
  \subfigure[]{
    \includegraphics[width = 0.46\linewidth]{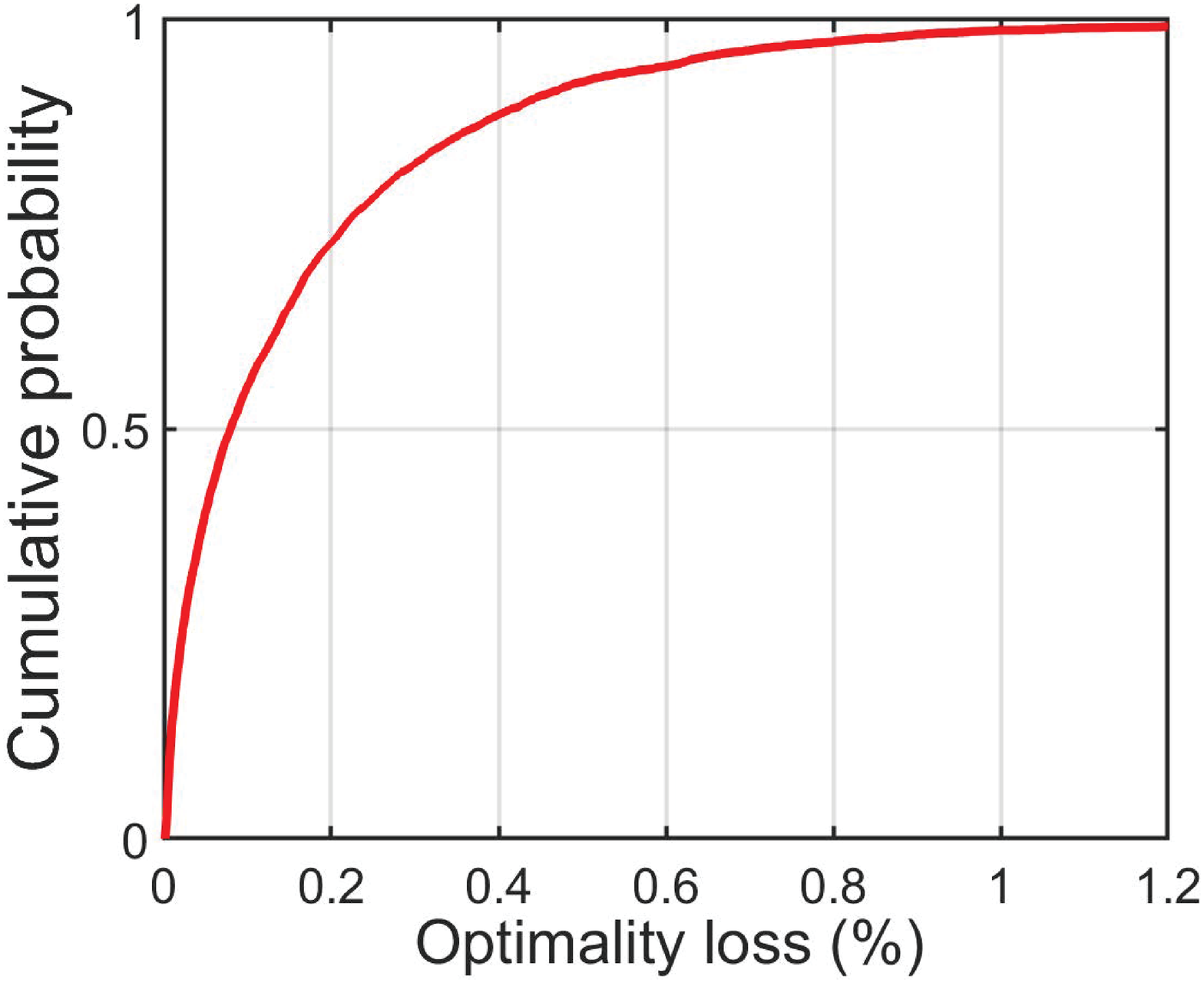}\label{figure6_b}
  }
  \caption{Empirical cumulative distribution of speedup and optimality loss for the IEEE Case118 under typical operating conditions.}
\end{figure}

\rev{\section{The statistical result of the speedup and the optimality loss for the IEEE Case118.} \label{apx:cdf}
We plot the empirical cumulative distribution of the speedup and the optimality loss for the IEEE Case118 in Fig.~\ref{figure6_a} and Fig.~\ref{figure6_b}, respectively. As compared to the optimal solution obtained by the Gurobi solver, \textsf{DeepOPF} achieves an average optimality loss less than 0.2\% with the maximum around 1.2\%. Meanwhile, as compared to the solving time used by the Gurobi solver, \textsf{DeepOPF} achieves an average speedup of $\times$281 with the maximum around $\times$320.}

\end{appendices}
\end{NoHyper}
\end{document}